\documentclass[a4paper,11pt]{article}
\pdfoutput=1 
\usepackage{jheppub} 
\usepackage[T1]{fontenc} 
\usepackage{indentfirst}
\usepackage{hhline}
\usepackage{tikz}
\usepackage{yfonts}
\newcommand{\zhuxi}{\textcolor{red}}
\newcommand{\haoyu}{\textcolor{blue}}
\newcommand{\ads}{AdS$_3$~}
\usepackage{knitting}
\knitnogrid
\usepackage{lineno}

\title{\boldmath Topological Entanglement Entropy in Euclidean AdS$_3$ via Surgery}

\author[{\tiny \textknit{t}}]{Zhu-Xi Luo}
\author[{\tiny \textknit{x}}]{and Hao-Yu Sun}

\affiliation[{\tiny \textknit{t}}]{Department of Physics and Astronomy, University of Utah,\\Salt Lake City, UT 84112, U.S.A.}
\affiliation[{\tiny \textknit{x}}]{Department of Physics, University of California, \\Berkeley, CA 94720, U.S.A.}

\emailAdd{zhuxi.luo@utah.edu}
\emailAdd{hkdavidsun@berkeley.edu}

\abstract{We calculate the topological entanglement entropy (TEE) in Euclidean asymptotic \ads spacetime using surgery. The treatment is intrinsically three-dimensional. In the BTZ black hole background, several different bipartitions are applied. For the bipartition along the horizon between two single-sided black holes, TEE is exactly the Bekenstein-Hawking entropy, which supports the ER=EPR conjecture in the Euclidean case. For other bipartitions, we derive an Entangling-Thermal relation for each single-sided black hole, which is of topological origin. After summing over genus-one classical geometries, we compute TEE in the high-temperature regime. In the case where $k=1$, we find that TEE is the same as that for the \emph{Moonshine double state}, given by the maximally-entangled superposition of $194$ types of ``anyons'' in the 3d bulk, labeled by the irreducible representations of the Monster group. We propose this as the bulk analogue of the thermofield double state in the Euclidean spacetime. 
Comparing the TEE between thermal \ads and BTZ solutions, we discuss the implication of TEE on the Hawking-Page transition in 3d.}

\begin{document} 
\maketitle

\section{Introduction}
\label{sec:intro}

Topological entanglement entropy (TEE), first introduced in condensed matter physics \cite{KitaevPreskill,LevinWen}, has been widely used to characterize topological phases. It is the constant subleading term (relative to the area-law term) in the entanglement entropy, only dependent on universal data of the corresponding topological phase. 

At low energy, a large class of topological phases can be effectively described using Chern-Simons gauge theory with a compact, simple, simply-connected gauge group. When this is the case, TEE can be found using surgery \cite{Fradkin} and replica trick \cite{Calabrese} by computing the partition function on certain $3$-manifolds. For compact gauge groups, TEE is expressed \cite{Fradkin} in terms of modular $S$ matrices of Wess-Zumino-Witten rational conformal field theory (RCFT) on a $2d$ compact Riemann surface, following the CS/WZW correspondence first described in geometric quantization by \cite{Witten1}.

In three-dimensional spacetime, gravity can be classically described by Chern-Simons gauge theory with a non-compact, possibly complex gauge group \cite{Witten1988}. Specifically, in Euclidean picture with a negative cosmological constant $\Lambda=-1/l^2<0$, in the first-order formulation of general relativity, the spin connection $\omega$ combines with the ``vierbein'' $e$ to make the holomorphic Chern-Simons gauge field $\omega+ e/l$ and anti-holomorphic gauge field $\omega- e/l$ of gauge group $SL(2,\mathbb{C})$, where $l$ is the AdS$_3$ curvature radius. The following questions thus arise naturally: 
is there a similar notion of TEE in 3d gravity? If so, can one compute the TEE for 3d gravity using surgery? Is the TEE related to modular $S$ matrices of a CFT living on the conformal boundary? In Ref. \cite{Verlinde}, the authors proposed that the Bekenstein-Hawking entropy of a BTZ black hole \cite{BTZ,BHTZ} in AdS$_3$ can be interpreted as TEE. The argument is supported by calculations in the dual CFT.  
Unfortunately it is still not clear what is the meaning of this entanglement entropy, i.e. what are the two subregions or components that are entangled together.

\begin{figure}[htbp]
\label{fig:a}
\centering
\begin{tikzpicture}
\draw[thick] (0,1) arc (90:-180:1 and 1);
\draw[thick] (-1,0) arc (-90:0:1 and 1);
\draw[thick,blue] (-1,0) arc (180:90:1 and 1);
\node at (-1,0.75) {\haoyu{a}};
\node at (1.2,-0.8) {1-a};
\node at (-0.5,0.5) {\haoyu{A}};
\node at (0.5,-0.2) {B};
\end{tikzpicture}
\caption{Bipartition of constant time slice of thermal \ads.}
\label{fig:tads-partition}
\end{figure}
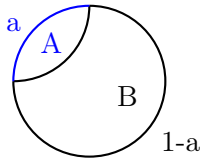

We are motivated by these questions to calculate TEE via 3d surgery in an Euclidean spacetime that is asymptotically AdS$_3$. In the case of thermal AdS$_3$, the constant time slice is a disk. We first bipartite this disk into two disks as shown in Fig. \ref{fig:tads-partition}, where $a$ denotes the ratio between the interval length on the boundary circle that is contained in subregion $A$ and the circumference of the full circle. After applying the replica trick, the glued manifold is a genus-$n$ handlebody. Using one-loop partition function on this handlebody \cite{Witten0706,YinXi,BinChen,MaloneyWitten,Giombi,DongXi}, we derive an explicit expression for TEE, which vanishes in the low-temperature limit.
Then we consider two disjoint thermal \ads and calculate the TEE between them, which turns out to be the thermal entropy of one thermal \ads. However, this does not mean any nontrivial entanglement between the two solid tori, and we support this argument by calculating the mutual information between them, which gives zero.

We also compute TEEs in an eternal BTZ background. In the Euclidean picture there is only one asymptotic region for the eternal BTZ black hole \cite{KrasnovEuclidean}, which corresponds to the gluing of the two asymptotic regions of the two single-sided black holes in the Lorentzian picture. We show that TEE between the two single-sided black holes is equal to the Bekenstein-Hawking entropy of one single-sided black hole. The mutual information between them does not vanish and again equals to the Bekenstein-Hawking entropy, which guarantees the explanation of the result as supporting the ER=EPR conjecture to be true \cite{MaldacenaTFD,Raamsdonk,ER=EPR}.

Focusing on one single-sided black hole, we then derive an Entangling-Thermal relation, stating
\begin{equation}
\lim_{\text{Area}(\bar{A})\rightarrow 0} [S(A)-S(\bar{A})] = S_{BTZ}^{\text{thermal}},
\end{equation}
where $A$ and $\bar{A}$ denotes the two complementary subregions. Quantities on both sides of this equation are intrinsically three-dimensional. The underlying physical reason of this relation is that, subregion $A$ wraps the non-contractible loop of the constant time slice, while its complement $\bar{A}$ does not. The difference between $S_A$ and $S_{\bar{A}}$ thus detects the effect of the non-contractible loop, which is exactly the outer horizon of the BTZ black hole. This relation is similar to but different from the thermal entropy relation \cite{Azeyanagi} derived from the Ryu-Takayanagi formula \cite{RT}, in that our result is topological and does not depend on geometrical details.

The full modular-invariant genus one partition function of three-dimensional pure gravity is a summation of classical geometries or gravitational instantons, which include both thermal \ads and the BTZ black hole. 
At high temperatures, the full partition function is dominated by the $SL(2,\mathbb{Z})$ family of black hole solutions, whereas the low-temperature solution is dominated by the thermal AdS$_3$. We compute TEE for the full partition function with a bipartition between the two single-sided black holes in the high temperature regime and again observe ER=EPR explicitly. When Chern-Simons level $k_R=k_L=l/16G=1$, after defining the quantum dimension data on the boundary Monster CFT, we see from the TEE calculation that the black hole geometries correspond to a topological phase in the bulk which contains a maximally-entangled superposition of $194$ types of ``anyons'', labeled by the irreducible representations of the Monster group. This state, dubbed as \emph{Moonshine double state}, has the similar property as the thermofield double state on the asymptotic boundary in that TEE between the anyon pairs is equal to the Bekenstein-Hawking entropy. 

The rest of the paper is organized as follows. In section \ref{sec:tool} we give a minimal introduction to the knowledge that facilitate the TEE calculation, including replica trick and Schottky uniformization. In section \ref{sec:tads} we show the calculation of TEE in thermal AdS$_3$, which amounts to the computation of the partition function on a genus $n$-handlebody. We also compute the TEE between two disjoint thermal \ads and show their mutual information vanishes. Section \ref{sec:btz} illustrates the TEE calculation for BTZ black holes for several different bipartitions. We discuss the relations with ER=EPR and show that mutual information between the two single-sided black holes is equal to the Bekenstein-Hawking entropy. We further propose an Entangling-Thermal relation for single-sided black holes. Then in section \ref{sec:whole} we demonstrate the TEE of the full modular-invariant partition function after summing over geometries and present the quantum dimension interpretation. The system is mapped to a superposition of $194$ types of anyons. Comments on the implication of TEE on the Hawking-Page transition and the outlook can be found in section \ref{sec:summary}.

\section{Review of Relevant Components}
\label{sec:tool}
In this section we will introduce basic concepts that are essential to understanding the rest of the paper.

\subsection{``Surgery'' and Replica Trick}

Surgery was originally invented by Milnor \cite{Milnor} to study and classify manifolds of dimension greater than three. 
In this work we use this concept in a broader sense, i.e. as a collection of techniques used to produce a new finite-dimensional manifold from an existing one in a controlled way. Specifically, it refers to cutting out parts of a manifold and replacing it by a part of another manifold, matching up along the cut.

\begin{figure}[htbp]
	\centering
	\begin{tikzpicture}[scale=0.4]
	\draw[thick] (-0.75,0) arc (180:10:0.75 and 3);
	\draw[thick] (-0.75,0) arc (-180:-10:0.75 and 3);
	\draw[thick] (0.7,0.5)--(3.5,0);
	\draw[thick] (0.7,-0.5)--(3.5,0);
	\draw[thick] (0,3)--(10,3);
	\draw[thick] (0,-3)--(10.75,-3);
	\draw[thick] (10,1) arc (-90:90:0.25 and 1);
	\draw[dashed,thick] (10,3) arc (90:270:0.25 and 1);
	\draw[thick] (10,1)--(8,1);
	\draw[thick] (10,1)--(11.5,1);
	\draw[thick] (11.5,-1) arc (-90:90:0.25 and 1);
	\draw[dashed,thick] (11.5,1) arc (90:270:0.25 and 1);
	\draw[thick] (11.5,-1)--(8,-1);
	\draw[thick]  (10.75,-1)--(11.5,-1);
	\draw[thick] (10.75,-3) arc (-90:90:0.25 and 1);
	\draw[dashed,thick] (10.75,-1) arc (90:270:0.25 and 1);
	\node[left] at (-1,0) {$3\beta$};
	\node at (10.7,2) {$\beta$};
	\node at (12.2,0) {$\beta$};
	\node at (11.5,-2) {$\beta$};
	\node at (2,-1) {$A$};
	\node at (7,-1) {$B$};
	\end{tikzpicture}~~~~
	\begin{tikzpicture}[scale=0.4]
	\draw[thick] (0,0) circle [x radius=0.75, y radius=3];
	\draw[thick] (0,3)--(10,3);
	\draw[thick] (0,-3)--(10.75,-3);
	\draw[thick] (10,1) arc (-90:90:0.25 and 1);
	\draw[dashed,thick] (10,3) arc (90:270:0.25 and 1);
	\draw[thick] (10,1)--(8,1);
	\draw[thick] (10,1)--(11.5,1);
	\draw[thick] (11.5,-1) arc (-90:90:0.25 and 1);
	\draw[dashed,thick] (11.5,1) arc (90:270:0.25 and 1);
	\draw[thick] (11.5,-1)--(8,-1);
	\draw[thick]  (10.75,-1)--(11.5,-1);
	\draw[thick] (10.75,-3) arc (-90:90:0.25 and 1);
	\draw[dashed,thick] (10.75,-1) arc (90:270:0.25 and 1);
	\node[left] at (-1,0) {$3\beta$};
	\node at (10.7,2) {$\beta$};
	\node at (12.2,0) {$\beta$};
	\node at (11.5,-2) {$\beta$};
	\node at (2,0) {$A$};
	\node at (7,0) {$B$};
	\end{tikzpicture}
	\caption{Left: Sketch of $\rho_A^3$. Right: Sketch of $\text{tr}\rho_A^3$}
	\label{fig:replica}
\end{figure}
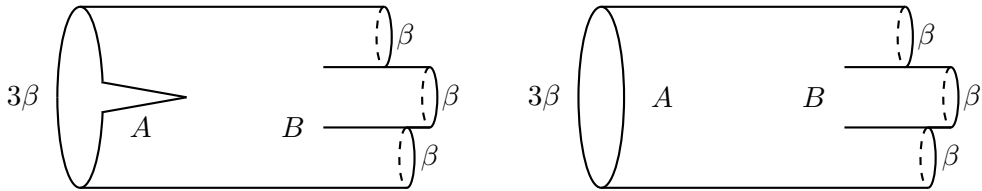

As a warm-up, we review the usage of surgery in the entanglement calculation of 2d CFT for a single interval at finite temperature $T=1/\beta$ \cite{Calabrese}.
The interval $A$ lies on an infinitely long line whose thermal density matrix is denoted as $\rho$. The reduced density matrix of subregion $A$ is then defined as $\rho_A=\text{tr}_{\bar{A}} \rho$, where the trace $\text{tr}_{\bar{A}}$ over the complement of $A$ only glues together points that are not in $A$, while an open cut is left along $A$. Entanglement entropy between $A$ and its complement $\bar{A}$ is then $S_A=
-\text{tr} \rho_A\ln\rho_A$. The matrix logarithm is generally hard to compute, so alternatively one applies the replica trick to obtain an equivalent expression, with proper normalization (so that the resultant quantity is $1$ when analytically continued to $n=1$):
\begin{equation}
S(A)=-\frac{d}{dn}\left(\frac{\text{tr} (\rho_A^n)}{(\text{tr}\rho_A)^n}\right)\Bigg|_{n=1}.
\end{equation}
Now the problem reduces to the computation of $\text{tr} (\rho_A^n)$. Using surgery, one can interpret it as the path integral on the glued 2-manifold \cite{CallanWilczek}. An example for $n=3$ is shown in Fig. \ref{fig:replica}, where the left panel sketches $\rho_A^3$, and the right panel is $\text{tr}(\rho_A^3)$. In this case with a finite temperature, $S_A$ is not necessarily equal to $S_{\bar{A}}$.

This operation can be extended to $3$-manifolds in a straightforward way, as shown in Ref. \cite{Fradkin}. The authors calculated examples where the constant time slices are closed surfaces and restricted to ground states, so that the $\beta$ cycle is infinitely long.

The constant time slices that we are interested in for Euclidean \ads are all open surfaces with asymptotic conformal boundaries, and the quantum states do not necessarily belong to the ground state Hilbert subspace. Details will be presented in sections \ref{sec:tads} and \ref{sec:btz}.

\subsection{Conformal Boundary and $\mathbb{H}^3/\Gamma$}

We now introduces the hyperbolic three-space $\mathbb{H}^3$ that describes the Euclidean AdS$_3$. It is the 3d analogue of hyperbolic plane, with standard Poincare-like metric
\begin{equation}\label{eq:metric}
ds^2=\frac{dy^2+dzd\bar{z}}{y^2},
\end{equation}
where $y>0$ and $z$ is a complex coordinate.

Any $3$-manifold $M$ having a genus $n$ Riemann surface $\Sigma_n$ as its conformal boundary that permits a complete metric of constant negative curvature can be constructed using Schottky uniformization. The idea is to represent the $3$-manifold $M$ as the quotient of $\mathbb{H}^3$ by a Kleinian group $\Gamma$ \cite{Thurston}, which is a discrete subgroup of $SL(2,\mathbb{C})$ as well as a discrete group of conformal automorphisms of $\Sigma_n$. 

The conformal boundary of $\mathbb{H}^3$ is a sphere at infinity, $S^2_{\infty}$, on which $\Gamma$ acts discretely, except for a \emph{limit set} of accumulation points of $\Gamma$ denoted by $\Lambda(\Gamma)$. The complement $\Omega(\Gamma)=S^2_{\infty}-\Lambda(\Gamma)$ is called the domain of discontinuity. Then the $3$-manifold $M$ has boundary $\Omega(\Gamma)/\Gamma$, a well-defined quotient.

In particular, when $M$ is a handlebody, $\Gamma$ reduces to a Schottky group, which is freely finitely generated by the loxodromic elements $\gamma_1,\dots,\gamma_n\in SL(2,\mathbb{C})$, that acts on $S^2_{\infty}$ as a fractional linear transformation. Among these generators, there are $3n-3$ independent complex parameters, which are coordinates on the Schottky space, a covering space of the complex moduli of the Riemann surface.

Each $\gamma\in\Gamma$ is completely characterized by its fixed points and its multiplier $q_\gamma$. An eigenvalue $q_\gamma$ is defined through the unique conjugation of $\gamma$ under $SL(2,\mathbb{C})$: $z\mapsto q_\gamma z$ with $|q_{\gamma}|<1$.
More explicitly, denoting $\eta,\,\xi$ as the fixed points of $\gamma$, one has
\begin{equation}
\frac{\gamma(z)-\eta}{\gamma(z)-\xi}=q_{\gamma}\frac{z-\eta}{z-\xi}.
\end{equation}

Within the Schottky group $\Gamma$, there are primitive conjugacy classes $\langle\gamma_1,\dots,\gamma_n\rangle$ of $\Gamma$, with ``primitive'' meaning that $\gamma$ is not a positive power of any other element in $\Gamma$.

\subsection{Solid Tori Classified as $M_{c,d}$}

The physical spacetimes we are concerned about in this paper are all solid tori, i.e. the $n=1$ case in the previous subsection. They have toroidal conformal boundaries, so the Schottky group actions is relatively simple.

After these topological constructions, we can further classify them into the $M_{c,d}$ family according to their geometries. This family first appeared in the discussion of classical gravitational instantons which dominate the path integral in Ref. \cite{SL2Z}, and is further explained in Refs. \cite{MaloneyWitten} and \cite{Tail}. 

In this case, $\Lambda(\Omega)$ composes of the north and south poles of $S^2_{\infty}$. Since solid tori have boundaries $T^2\cong\Omega(\Gamma)/\Gamma$, $\pi_1(\Omega(\Gamma))$ must be a subgroup of $\pi_1(T^2)$, so $\pi_1(\Omega(\Gamma))$ can only be isomorphic to $\mathbb{Z}\oplus\mathbb{Z}$, $\mathbb{Z}$, or the trivial group. When $\pi_1(\Omega(\Gamma))=\mathbb{Z}\oplus\mathbb{Z}$, $\Omega(\Gamma)$ has to be a Riemann surface of genus 1, which cannot be isomorphic to an open subset of $S^2_{\infty}$. When $\pi_1(\Omega(\Gamma))$ is trivial, $\Omega(\Gamma)$ is a simply-connected universal cover of $T^2$, so that $\Gamma$ has to be $\mathbb{Z}\oplus\mathbb{Z}$. It is easily seen from \eqref{eq:metric} that if $\Gamma\cong\mathbb{Z}\oplus\mathbb{Z}$, then although $\mathbb{H}^3/(\mathbb{Z}\oplus\mathbb{Z})$ has a toroidal boundary at $y=0$, there is a cusp at $y\rightarrow\infty$, whose sub-Plackian length scale invalidates semi-classical treatments. 

The only possibility is thus $\pi_1(\Omega(\Gamma))=\mathbb{Z}$, where $\Gamma$ can be either $\mathbb{Z}$ or $\mathbb{Z}\oplus\mathbb{Z}_n$. The latter yields $M$ to be a $\mathbb{Z}_n$-orbifold, indicating the existence of massive particles, which are not allowed in pure gravity. To avoid undesirable geometries such as cusps and orbifolds in the contributions to path integral \cite{Giombi,MaloneyWitten}, we restrict our Schottky group to be $\Gamma\cong\mathbb{Z}$, generated by the matrix
\begin{equation}
\label{eq:generator}
W=\left(\begin{matrix}
q & 0\\
0 & ~q^{-1}\\
\end{matrix}\right)
\end{equation}
where $|q|<1$. 

The boundary torus is thus obtained by quotiening the complex $z$-plane without the origin by $\mathbb{Z}$. Redefine $z=e^{2\pi i\omega}$, so $\omega$ is defined up to $\omega\rightarrow\omega+1$, and $W$ acts by $\omega\rightarrow\omega+\ln q/2\pi i$. Hence, the complex modulus of the torus is $\tau\equiv\ln q/2\pi i$, defined up to a $PSL(2,\mathbb{Z})$ M\"{o}bius transformation $\tau\sim(a\tau+b)/(c\tau+d)$, where integers $a,b,c,d$ satisfy $ad-bc=1$. 


When constructing a solid torus from its boundary torus, $\tau$ is defined only up to $\tau\sim\tau+\mathbb{Z}$ by a choice of solid filling, completely determined by the pair $(c,d)$ of relatively prime integers. This is because the flip of sign $(a,b,c,d)\rightarrow(-a,-b,-c,-d)$ does not affect $q$, and once $(c,d)$ are given, $(a,b)$ can be uniquely determined by $ad-bc=1$ up to a shift $(a,b)\rightarrow(a,b)+t(c,d),\, t\in\mathbb{Z}$ which leaves $q$ unaffected. We call these solid tori $M_{c,d}$'s, and any $M_{c,d}$ can be obtained from $M_{0,1}$ via a modular transformation on $\tau$. Physically, $M_{0,1}$ is the Euclidean thermal AdS$_3$ and $M_{1,0}$ is the traditional Euclidean BTZ black hole obtained from Wick rotating the original metric in \cite{BTZ}. Excluding $M_{0,1}$, $M_{c,d}$'s are collectively called the $SL(2,\mathbb{Z})$ family of Euclidean black holes, to be discussed in section \ref{sec:whole}.

\section{Thermal \ads}
\label{sec:tads}

The Euclidean thermal \ads has the topology of a solid torus $M_{0,1}$, whose non-contractible loop is parametrized by the Euclidean time. 
The constant time slice is thus a disk $D^2$ with a boundary $S^1$, perpendicular to the non-contractible loop. 

\subsection{Bipartition into Two Disks}

\begin{figure}[htbp]
	\centering
	\includegraphics[scale=0.18]{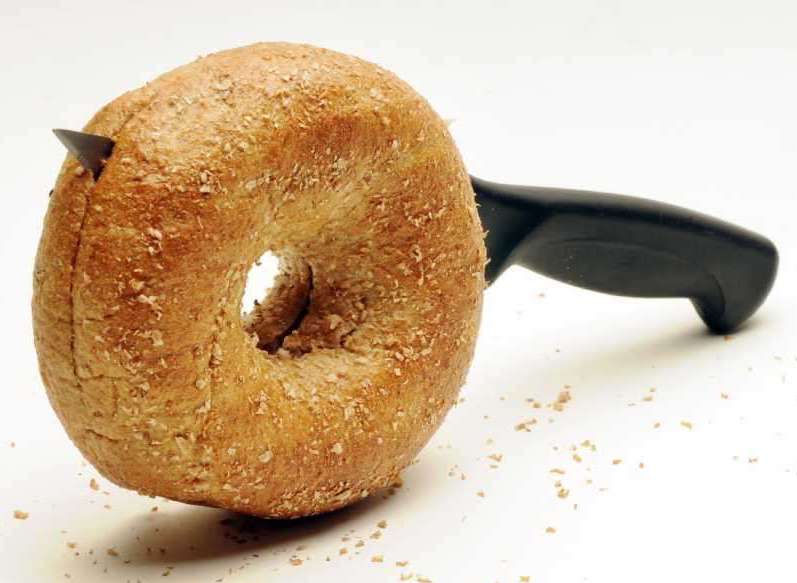}
	\hskip 1cm
	\begin{tikzpicture}[scale=0.8]
	\draw[thick] (0,0) arc (230:-50:1);
	\draw[thick] (0,0) arc (50:330:1);
	\draw[thick] (0.244,-1.27) arc (140:400:1);
	\draw[thick] (-1.15,-0.7) arc (-170:-10:0.5 and 0.25);
    \draw[thick] (-0.95,-0.85) arc (180:0:0.3 and 0.15);
    \draw[thick] (0.5,-1.9) arc (-170:-10:0.5 and 0.25);
    \draw[thick] (0.7,-2.05) arc (180:0:0.3 and 0.15);
    \draw[thick] (0.15,0.9) arc (-170:-10:0.5 and 0.25);
    \draw[thick] (0.35,0.75) arc (180:0:0.3 and 0.15);
    \node at (2,-0.5) {$\cdots$};
	\end{tikzpicture}
	\caption{\textbf{Left}: bipartition of the thermal AdS$_3$.
\textbf{Right}: the glued 3-manifold is a flat bouquet-like $n$-handlebody.}
	\label{fig:tads}
\end{figure}
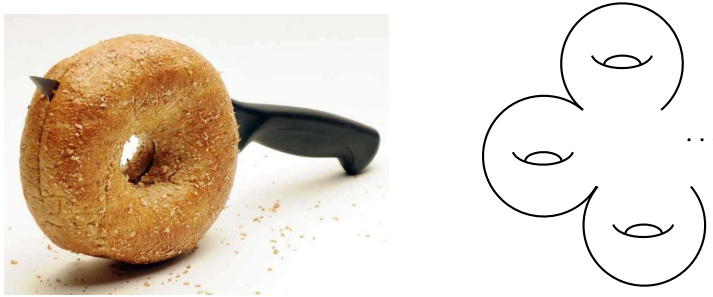

We bipartite the disk into upper and lower subregions $A$ and $B$, both having the topology of a disk. The solid torus is then turned into a sliced bagel as in Fig. \ref{fig:tads}. Boundary of each subregion contains an interval lying on the $S^1$. In the following we will denote the ratio between the length of one interval and the circumference of the boundary $S^1$ to be $a$, satisfying $0\leq a \leq 1$. Except for the symmetric case where $a=1/2$ and the two subregions are equivalent, generally $S_A\neq S_B$.

As introduced in section \ref{sec:tool}, one then glues each of $n$ copies of subregion $B$'s separately while gluing the $n$ copies of subregion $A$'s together. The resultant 3-manifold is an $n$-handlebody, which is a filled genus-$n$ Riemann surface, shown in Fig. \ref{fig:tads}. (In the special case of $n=1$, the handlebody reduces to a solid torus.)

With a proper normalization, the entanglement entropy corresponding to subregion $A$ is then
\begin{equation}
\label{eq:tads-replica}
S_{TAdS}=-\frac{d}{dn}\left(\frac{Z(n\text{-handlebody})}{Z(1\text{-handlebody})^n}\right)\Bigg|_{n=1}.
\end{equation}

Contribution to the path integral around a classical saddle point for an $n$-handlebody takes the form
\begin{equation}
Z(n)=\exp \left[ k S_0(n) + \sum_i k^{-i+1} S_i(n) \right],
\end{equation}
where $k^{-i+1}S_i(n)$ is the $i$-loop free energy of boundary graviton excitations. At tree level ($i=0$), $Z_{tree}(n\text{-handlebody})$ can be derived assuming the dual CFT is an extremal CFT \cite{YinXi}\footnote{This partition function is motivated by the Liouville action of a single free boson on a handlebody, and is conjectured in \cite{YinXi} as a weight $12k$ modular form to avoid singularities of special functions.},
\begin{equation}
\label{eq:tads-tree}
Z_{tree}(n)=\prod_{\gamma~\text{prim.}}\prod_{m=1}^\infty |1-q_\gamma^m| ^{24k},
\end{equation}
with the product running over primitive conjugacy classes of $\gamma$, $q_\gamma$ being the multiplier of $\gamma$ introduced in section \ref{sec:tool}, and $k=l/16G$. 

In general the two products are hard to evaluate. However, in the low-temperature regime when thermal \ads dominates, the leading contribution to the infinite product over $m$ comes from $m=1$. Furthermore, the product over $\gamma$ is dominated by a single-letter contribution \cite{DongXi,Das}, $\prod\limits_{\gamma~\text{prim.}} |1-q_\gamma| \approx |1-q_1|^{2n}$. Combining these, we obtain
\begin{equation}
Z_{tree}(n)\approx \prod_{\gamma~\text{prim.}} |1-q_1|^{24k}=|1-q_1|^{48nk},
\end{equation}
with $q_1$ a function of $n$ and $a$, having the form
\begin{equation}
\label{eq:tads-q1}
q_1=\frac{\sin^2(\pi a)}{n^2\sin^2(\pi a/n)}e^{-2\pi\beta}.
\end{equation}

At one-loop ($i=1$) level, the general expression for $Z_{loop}(n\text{-handlebody})$ can be derived from either the boundary extremal CFT \cite{YinXi,BinChen} or the bulk heat kernel method \cite{Giombi}. They both depend on the Schottky parametrization of the boundary genus $n$-Riemann surface. The result is
\begin{equation}\label{eq:tads-loop}
Z_{loop}(n)=\prod_{\gamma~\text{prim.}} \prod_{m=2}^\infty \frac{1}{|1-q_\gamma^m|}\approx \frac{1}{|1-q_1^2|^{2n}},
\end{equation}
in the low-temperature regime $q_1\ll 1$. Plugging $Z(n\text{-handlebody})=Z_{tree}(n)Z_{loop}(n)$ into \eqref{eq:tads-replica}, we obtain
\begin{equation}
\label{eq:ads3}
S_{TAdS}(a)\approx\left[96k e^{-2\pi\beta}+(96k-8) e^{-4\pi\beta} + O(e^{-6\pi\beta})\right]\left(\pi a\cot(\pi a)-1\right).
\end{equation}
The terms containing $k$ come from tree-level, while others are one-loop contributions. The entire expression approaches to zero very fast in the low-temperature regime $\beta\rightarrow\infty$ for any $k$. The dependence of the above result on $a$ distinguishes itself from the original definition \cite{KitaevPreskill, LevinWen} of TEE, which is a universal constant. We note that $a$ enters as the boundary condition on the constant time slice, and has nothing to do with the leading area-law term in usual expressions of entanglement entropies.

When subregion $A$ is ``nothing'', i.e. $a\rightarrow 0$, $\pi a\cot(\pi a)\rightarrow 1$, thus the TEE between $S_A$ vanishes. When $A$ is instead ``everything'', i.e. $a\rightarrow 1$, $\pi a \cot(\pi a)\rightarrow -\infty$, balanced by the smaller $e^{-2\pi\beta}\ll 1$ at low temperatures. We observe that apart from the $a\rightarrow 0$ case, the TEE for thermal \ads is always negative. Another important case is when $a=1/2$ so that the two subregions are symmetric. In this case we have 
\begin{equation}
\label{eq:tads-symmetric}
S_{TAdS}\left(a=\frac{1}{2}\right)\approx-\left[96k e^{-2\pi\beta}+(96k-8) e^{-4\pi\beta} + O(e^{-6\pi\beta})\right].
\end{equation}

\subsection{Two Disjoint Thermal \ads}

Now we take two non-interacting thermal \ads as the whole system, represented by two disjoint solid tori $M_{0,1}$.  There are two non-interacting, non-entangled, identical CFTs living on their asymptotic boundaries. One would naively expect the TEE between these two solid tori to be zero, which is not really the case. To calculate the entanglement entropy between these two solid tori, one can simply use
\begin{equation}
    \label{eq:tads-epr}
    S_{TAdS}=-\frac{d}{dn}\left(\frac{Z_{0,1}(n\tau)Z_{0,1}(\tau)^n}{Z_{0,1}(\tau)^{2n}}\right)\Bigg|_{n=1}.
\end{equation}
We have used the shorthand notation $Z_{0,1}(\tau)=Z_{0,1}(\tau,\bar{\tau})$ to take into account both holomorphic and anti-holomorphic sectors. The partition function $Z_{0,1}(n\tau)$ comes from gluing $n$ copies of solid torus $A$, which is a new solid torus with modular parameter $n\tau$. 

Meanwhile, $Z_{0,1}(\tau)^n$ comes from gluing individually the $n$ copies of solid torus $B$. We can simply multiply the contributions from $A$ and $B$ together because they are disjoint. Then we can plug these into the expression for the solid torus partition function, i.e. the $1$-handlebody result from \eqref{eq:tads-tree} and \eqref{eq:tads-q1},
\begin{equation}
    Z_{0,1}(\tau)=|q|^{2k}\prod_{m=2}^\infty |1-q^m|^{-2}.
\end{equation}
In the low temperatures, we can approximate $q=e^{2\pi i\tau}=e^{-2\pi\beta}$ as a small number and thus at leading order $Z_{0,1}(\tau)\approx q^{-2k}(1-q^2)^{-2}$.

After straightforward calculations we obtain 
\begin{equation}
S_{TAdS}\approx 2(1+4\pi\beta)e^{-4\pi\beta}.
\end{equation}
This contains only the loop contribution, i.e. the semi-classical result is zero. For comparison, we also calculate the canonical ensemble thermal entropy of a single thermal \ads at temperature $\beta^{-1}$: $S_{TAdS}^{\text{thermal}}=\ln Z(1\text{-handlebody})-\beta Z(1\text{-handlebody})^{-1} \frac{\partial Z(1\text{-handlebody})}{\partial \beta}.$
It has the low-temperature form
\begin{equation}
S_{TAdS}^{\text{thermal}}\approx 2(1+4\pi\beta)e^{-4\pi\beta},
\end{equation}
which again solely comes from loop contributions. We immediately observe that the thermal entropy of a single thermal \ads is the same as the TEE between two independent thermal \ads. 

This does {\it not} imply that there are nontrivial topological entanglement between the two copies of thermal AdS$_3$, but simply reveals the insufficiency of using entanglement entropy as an entanglement measure at finite temperatures. For example, consider two general subsystems $A$ and $B$ with thermal density matrices $\rho_A$ and $\rho_B$ and combine them into a separable system,
\begin{equation}
    \rho=\rho_A\otimes \rho_B.
\end{equation}
These two subregions are thus obviously non-entangled. But if one attempts to calculate the entanglement entropy between $A$ and $B$ by tracing over $B$, one can still get an arbitrary result depending on the details of $\rho_A$. If we choose $\rho_A=|\psi\rangle \langle \psi|$ where $|\psi\rangle$ is some pure state, then the entanglement entropy will be zero. If instead we choose $\rho_A=\frac{1}{\dim (\mathcal{H}_A)}\mathbf{1}$ as the proper normalized identity matrix, then the entanglement entropy will be $\ln(\dim (\mathcal{H}_A))$. So depending on the choice of $\rho_A$, one can obtain any value of the entanglement entropy between these minimum and maximum values. This shortcoming is due to the fact that now the entanglement entropy calculation involves undesired classical correlations in mixed states.

To address this issue, we look at the topological mutual information between the two solid tori,
\begin{equation}
I(A,B)=S(A)+S(B)-S(A\cup B),
\end{equation}
so that the thermal correlations can be canceled. Following similar replica trick calculations, one easily obtain $S(A\cup B)=2S(A)=2S(B)$, thus the mutual information vanishes and there exists no nontrivial topological entanglement between the two disjoint thermal AdS$_3$. We will observe in the next section that this statement no longer holds true for an eternal BTZ black hole.

\section{BTZ Black Hole}
\label{sec:btz}
We will explore in this section the topological entanglement in the bulk of Euclidean BTZ black hole.

\subsection{BTZ Geometry}
It has been speculated for a long time that the 3d gravity is rather trivial because there is no gravitational wave besides local fluctuations. However in 1992, authors of \cite{BTZ} proposed a new type of AdS-Schwarzschild black hole with Lorentzian metric
\begin{equation}
\label{eq:btz}
ds_L^2=-N_L^2dt_L^2+N_L^{-2}dr^2+r^2(d\phi+N_L^\phi dt)^2,
\end{equation}
where the lapse and shift functions have the form $N_L^2=-8G M_L +\frac{r^2}{l^2}+\frac{16G^2J_L^2}{r^2},~~N_L^\phi=-\frac{4G J_L}{r^2}.$
$G$ is the three-dimensional Newton constant, $l$ the curvature radius of AdS$_3$, and $M$, $J_L$ are the mass and angular momentum of the black hole, respectively. The outer and inner horizons are defined by
\begin{equation}
r_\pm^2=4GM_Ll^2\left(1\pm \sqrt{1-\frac{J_L^2}{M_L^2l^2}}\right). 
\end{equation}

Let $t_L=i t$ and $J_L=iJ$, and we do the Wick rotation to get
\begin{equation}
\label{eq:wick}
ds^2=N^2dt^2+N^{-2}dr^2+r^2(d\phi+N^\phi dt)^2,
\end{equation}
with $N^2=-8G M +\frac{r^2}{l^2}-\frac{16G^2J^2}{r^2},~~N^\phi(r)=-\frac{4G J}{r^2}$. The horizons are now given by
\begin{equation}
r_\pm^2=4GMl^2\left(1\pm \sqrt{1+\frac{J^2}{M^2l^2}}\right).
\end{equation}
The Euclidean BTZ black hole is locally isometric to the hyperbolic three-space $\mathbb{H}^3$ and is globally described by $\mathbb{H}^3/\Gamma$ with $\Gamma\cong\mathbb{Z}$. The topology is a solid torus, and one can make it explicit by doing the following coordinate transformations \cite{CoordianteTransformation}
\begin{equation}
\begin{split}
\label{eq:spherical}
x &=\sqrt{\frac{r^2-r_+^2}{r^2-r_-^2}}\cos\left(\frac{r_+}{l^2}t+\frac{|r_-|}{l}\phi\right)\exp\left(\frac{r_+}{l}\phi-\frac{|r_-|}{l^2}t\right),\\
y &=\sqrt{\frac{r^2-r_+^2}{r^2-r_-^2}}\sin\left(\frac{r_+}{l^2}t+\frac{|r_-|}{l}\phi\right)\exp\left(\frac{r_+}{l}\phi-\frac{|r_-|}{l^2}t\right),\\
z &=\sqrt{\frac{r_+^2-r_-^2}{r^2-r_-^2}}\exp\left(\frac{r_+}{l}\phi-\frac{|r_-|}{l^2}t\right)>0.\\
\end{split}
\end{equation}
They bring the metric \eqref{eq:wick} to the upper half-space $\mathbb{H}^3$ with $z>0$. Further changing to the spherical coordinates $(x,y,z)=(R\cos\theta\cos\chi,R\sin\theta\cos\chi,R\sin\chi)$, we finally arrive at
\begin{equation}
\label{eq:fundamental}
ds^2=\frac{l^2}{\sin^2\chi}\left(\frac{dR^2}{R^2}+\cos^2\chi d\theta^2+d\chi^2\right).
\end{equation}

To ensure that the above coordinate transformation is non-singular (contains no conical singularities) at the $z$ axis $r=r_+$, we must require periodicity in the arguments of the trigonometric functions. That is, we must identify
\begin{equation}
\frac{1}{2\pi l} \left(\phi,t\right)\sim \frac{1}{2\pi l}(\phi+\Phi,t+\beta), 
\end{equation}
where $\Phi=\frac{|r_-|}{r_+^2-r_-^2},\,\beta=\frac{r_+ l}{r_+^2-r_-^2}.$ We recombine the real pair $(\Phi,\beta)$ into a single complex variable 
\begin{equation}
\label{eq:tau}
\tau=\Phi+i\beta,
\end{equation}
which is the complex modular parameter of the boundary torus. In terms of metric \eqref{eq:fundamental}, this corresponds to the global identifications
\begin{equation}
(R,\theta,\chi)\sim\left(Re^{2\pi r_+/l},\theta+\frac{2\pi|r_-|}{l},\chi\right).
\end{equation}

\begin{figure}[htbp]
	\centering
	\begin{tikzpicture}[scale=0.4]
	\draw[gray,->] (0,0)--(5,0);
	\draw[gray,->] (0,0)--(0,6);
	\draw[gray,->] (0,0)--(-2.5,-2.5);
	\draw[thick] (-4,0) arc (-180:0:4 and 2);
	\draw[thick] (4,0) arc (0:180:4 and 2);
	\draw[thick] (-2,0) arc (-180:0:2 and 1);
	\draw[dashed,thick] (2,0) arc (0:180:2 and 1);
	\draw[thick] (4,0) arc (0:180:4 and 5);
	\draw[thick] (2,0) arc (0:180:2 and 3);
	\node at (5,-0.5) {$y$};
	\node at (-3,-2.5) {$x$};
	\node at (0.5,6) {$z$};
	\end{tikzpicture}
	\hskip 1cm
	\begin{tikzpicture}[scale=0.5]
	\draw[thick] (0,0) circle [x radius=4, y radius=2.5];
	\draw[thick] (-1.5,0.41) arc (-170:-10:1.5 and 0.9375);
	\draw[thick] (-1,-0.15) arc (180:0:1 and 0.625);
	\draw[dashed,blue,thick] (0,0) circle [x radius=2.8, y radius=1.5];
	\draw[white] (0,-3.5)--(3,-3.5);
	\end{tikzpicture}
	\caption{(Color online) \textbf{Left}: The spherical coordinates on $\mathbb{H}^3$, which converts the original Schwarzschild metric \eqref{eq:btz} of BTZ black hole into the right picture. \textbf{Right}: Topology of the Euclidean BTZ black hole is a solid torus. Horizon is the blue dashed line threading the central cord of the solid torus. The Euclidean time runs in the meridian direction.}
	\label{fig:BTZ}
\end{figure}
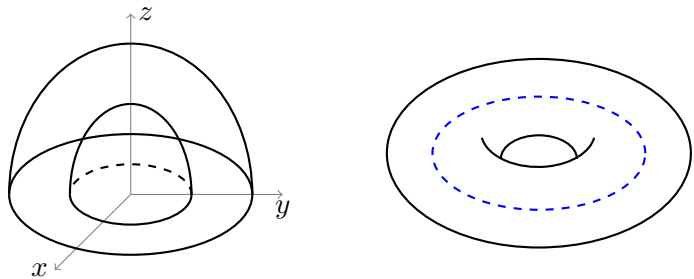

A fundamental region for \eqref{eq:fundamental} is the filling of the slice between inner and outer hemispheres centered at the origin having radii $R=1$ and $R=e^{2\pi r_+/l}$ respectively, with an opening $2\pi|r_-|/l$ or $2\pi$ (if $r_-=0$) in azimuthal angle, as shown by Fig. \ref{fig:BTZ}, and two hemispheres are identified along the radial lines with a twist of angle $2\pi|r_-|/l$ or $2\pi$. Hence, the segment on $z$-axis between two hemispheres corresponding to the outer horizon, and is mapped to the central cord of solid torus at $\chi=\pi/2$ (the boundary torus is at $\chi=0$). 

For convenience, in the rest of the paper, unless stated otherwise, we only focus on non-rotating Euclidean BTZ black hole, so that $\tau$ is pure imaginary and $r_-=0$.

\subsection{TEE between Two One-Sided Black Holes and Mutual Information}
\label{subsec:btz-erepr}

Following Refs. \cite{MaldacenaTFD,Raamsdonk,ER=EPR}, an eternal Lorentzian AdS black hole has two asymptotic regions and can be viewed as two black holes connected through a non-transversable wormhole. It is also suggested from the dual CFT perspective that the entanglement entropy between the CFTs living on the two asymptotic boundaries is equal to the thermal entropy of one CFT. Motivated by this, we are interested in calculating the TEE between the two single-sided black holes in the bulk.
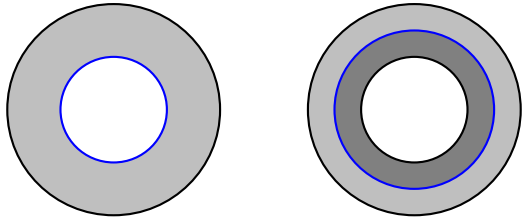
\begin{figure}[htbp]
	\centering
	\begin{tikzpicture}[scale=0.7]
	\filldraw[fill=lightgray,  draw=black, thick] (0,0) circle [radius=2];
	\filldraw[fill=white, draw=blue, thick] (0,0) circle [radius=1];
	\end{tikzpicture}
	\hskip 1cm
	\begin{tikzpicture}[scale=0.7]
	\filldraw[fill=lightgray, draw=black, thick] (0,0) circle [radius=2];
	\filldraw[fill=gray,  draw=blue, thick] (0,0) circle [radius=1.5];
	\filldraw[fill=white,  thick] (0,0) circle [radius=1];
	\end{tikzpicture}
	\caption{(Color online) \textbf{Left}: Constant time slice of each single-sided BTZ black hole is an annulus. The inner boundary in blue denotes the horizon. Time evolution of this slice corresponds to rotating angle $\pi$ around the inner blue boundary. \textbf{Right}: Gluing the constant time slices of single-sided black hole $R$ (light grey) and $L$ (dark grey) along the horizon (blue line) in the middle.}
	\label{fig:ConstantBTZ}
\end{figure}

However, for the Euclidean BTZ black hole \eqref{eq:wick} and \eqref{eq:fundamental}, the metrics only cover the spacetime outside the horizon of one single-sided black hole. Everything inside the horizon is hidden, including the other single-sided black hole. In order to make the computation of TEE between two single-sided black holes possible, we take an alternative view of the solid torus $M_{1,0}$, as in Fig. \ref{fig:ConstantBTZ}. In the left panel, we sketch the constant time slice of the right single-sided black hole, call it $R$. It is the constant $\theta$ slice in metric \eqref{eq:fundamental} with an annulus topology, whose inner boundary is identified with the horizon. In the right panel, we glue the two constant time slices for black holes $L$ and $R$ along the horizon. Then comes the most important step: we fold the annulus of black hole $L$ along the horizon, so that it coincides with the annulus of black hole $R$. To obtain the full spacetime geometry, one rotates the constant time slice of $L$ about the horizon \emph{counterclockwise} by $\pi$, while rotating the constant time slice of $R$ about the horizon \emph{clockwise} by $\pi$. Namely, the two annuli meet twice: one at angle $0$, the other at $\pi$. The resultant manifold is a solid torus, same as the $M_{1,0}$ introduced before. Hence one can view this solid torus either as one single-sided black hole $R$ with modular parameter $\tau=i\beta$, or as two single-sided black holes $L$ and $R$, each contributing $\tau'=i\beta/2$. 

It might concern some readers that the CFTs living on the asymptotic boundaries of $L$ and $R$ in the Lorentzian picture are now glued together. We note that this is a feature of the Euclidean picture: due to the different direction of evolutions, we have CFT$_L(t)=$CFT$_R(-t)$. At $t=0$, these obviously coincide. Then at $t=\beta/2$, this gives CFT$_L(t=\beta/2)=$CFT$_R(t=-\beta/2)$. Using the fact that in the Euclidean picture we have $-\beta/2=-\beta=2+\beta=\beta/2$, we arrive at CFT$_L(t=\beta/2)=$CFT$_R(t=\beta/2)$, thus they coincide again and the two CFTs are glued together. This is consistent with the fact that in the Euclidean signature, there should only be one asymptotic region, as shown in \cite{KrasnovEuclidean}.

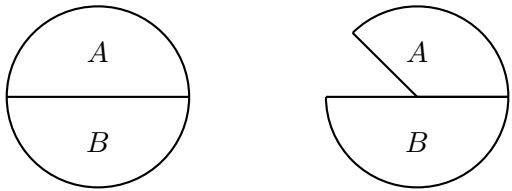
\begin{figure}[htbp]
\centering
\begin{tikzpicture}[scale=0.6]
\draw[thick] (0,0) circle [radius=2];
\draw[thick] (-2,0)--(2,0);
\node at (0,1) {$A$};
\node at (0,-1) {$B$};
\draw[thick] (5.586,1.414) arc (135:-180:2 and 2);
\draw[thick] (5,0)--(9,0);
\draw[thick] (7,0)--(5.586,1.414);
\draw[thick] (7,0)--(9,0);
\node at (7,1) {$A$};
\node at (7,-1) {$B$};
\end{tikzpicture}
\caption{The disk perpendicular to the horizon, which pierces the center of the disk. \textbf{Left:} Here, parts $A$ and $B$ in spacetime are respectively formed by rotating both spatial subregions $A$ and $B$ by $\pm\pi$. \textbf{Right:} The graphical representation of $\rho_A$, with a wedge missing in spacetime subregion $A$.}
\label{fig:beta-p}
\end{figure}

Now we can calculate the TEE between the constant time slices of $L$ and $R$, which we denote as $A$ and $B$. Importantly, since in general the result can be time dependent, we specify the cut to be done at $t=0$. Shown in the left panel of Fig. \ref{fig:beta-p}, each subregion contributes $\tau'$ to the modular parameter of the solid torus. We sketch one copy of $\rho_A$ in the right panel.

\begin{figure}[htbp]
	\centering
	\begin{tikzpicture}[scale=0.6]
	\draw[thick] (0,0) circle [x radius=4, y radius=2.5];
	\draw[thick] (-1.5,0.41) arc (-170:-10:1.5 and 0.9375);
	\draw[thick] (-1,-0.15) arc (180:0:1 and 0.625);
	\draw[blue,thick,dashed] (0,0) circle [x radius=2.8, y radius=1.5];
	\draw[blue,thick] (-2.12,-0.953) arc (-145:-35:2.62 and 1.28);
	\draw[thick] (-2.1,-1)--(-2.3,-0.4);
	\draw[thick] (2.1,-1)--(2.3,-0.4);
	\draw[thick] (-2.3,-0.4).. controls (0,-1.2)..(2.3,-0.4);
	\draw[thick] (-2.1,-1)--(-2.5,-1.6);
	\draw[thick] (2.1,-1)--(2.3,-1.6);
	\draw[thick] (-2.5,-1.6)..controls(0,-2)..(2.3,-1.6);
	\draw[thick] (9,0) circle [radius=2];
	\draw[thick] (9,0) -- (11,0);
	\draw[thick] (9,0)--(9,2);
	\draw[thick] (9,0)--(10,1.73205);
	\draw[thick] (9,0)--(10.73205,-1);
	\draw[thick] (9,0)--(9,-2);
	\node at (8.2,-0) {$\cdots$};
	\node at (10.2,0.75) {{\small $B_1$}};
	\node at (10.4,-0.4) {{\small $A_2$}};
	\node at (9.4,1.5) {{\small $A_1$}};
	\node at (9.6,-1) {{\small $B_2$}};
	\end{tikzpicture}
	\caption{(Color online) \textbf{Left}: front view of the pictorial representation of $\rho_A$. Notice that the cutaway wedge runs along the longitude (non-contractible loop) of the solid torus, with its vertex on the horizon. \textbf{Right}: Graphical representation of $\text{tr}\rho_A^n$. The disk is perpendicular to the horizon.}
	\label{fig:BTZsurgery}
\end{figure}
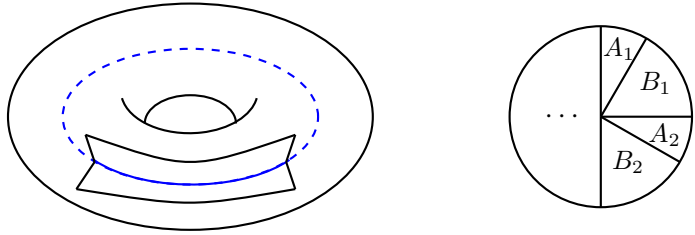

To find $S(A)$, we need to calculate the partition function of the 3-manifold that correspond to $\text{tr}\rho_A^n$. We first enlarge the missing wedge in the right panel of Fig. \ref{fig:beta-p} and shrink the size of $A$, $B$. To add the second copy of $\rho_A$, one should glue $A_1$ to $B_2$, with $B_2$ glued with $A_2$, as shown in Fig. \ref{fig:BTZsurgery}. Note that this differs from the usual way of doing replica tricks, where $A_1$ is always glued to $A_2$. This is again a result of the opposite directions of time evolutions for $L$ and $R$: the $B$ spatial slice at $t=\beta/2$ should always be identified with the $A$ spatial slice at $t=\beta/2$. One can then follow this procedure and glue $n$-copies of $\rho_A$.

The resultant 3-manifold is a solid torus with modular parameter $2n\tau'$, since each copy of $A$ contributes $\tau'$ and the same goes for $B$. Replica trick then gives
\begin{equation}
\label{eq:replica}
    S_{BTZ}(A)=-\frac{d}{dn}\left( \frac{Z_{1,0}(2n\tau')}{Z_{1,0}(2\tau')^n} \right)\Bigg|_{n=1}.
\end{equation}
Partition function $Z_{1,0}(\tau)$ can be obtained from that of the thermal \ads by a modular transformation $\tau\rightarrow -1/\tau$,
\begin{equation}
\label{eq:btz-partition}
Z_{1,0}(\tau)=|q_-|^{-2k}\prod_{m=2}^\infty \frac{1}{|1-q_-^m|^2},
\end{equation}
where we have defined $q_-\equiv e^{-2\pi i/\tau}=e^{-2\pi/\beta}$. In the high-temperature regime $\beta\ll 1$, the above reduces to
$Z_{1,0}(\tau)\approx e^{4\pi k/\beta} \left(1-e^{-4\pi/\beta}\right)^{-2}.$ Substituting it into \eqref{eq:replica}, one obtains at leading order
\begin{equation}
\label{eq:btz-thermal-A}
    S_{BTZ}(A) =\frac{8\pi k}{\beta}-2 e^{-4\pi/\beta} \left(\frac{4\pi}{\beta}-1\right)+O(e^{-6\pi/\beta}).
\end{equation}
where the first term comes from tree level and is identified with the Bekenstein-Hawking entropy. The above expression matches with the thermal entropy of one single-sided black hole at one-loop,
\begin{equation}
    S_{BTZ}^{\text{thermal}}(A)=\ln Z_{1,0}(\tau)-\beta Z_{1,0}(\tau)^{-1} \frac{\partial Z_{1,0}(\tau)}{\partial \beta}=S_{BTZ}(A).
\end{equation}
Remarkably, this equation holds true regardless of $Z_{1,0}(\tau)$'s specific form.

It might be confusing at first that the Bekenstein-Hawking entropy, usually viewed as an area-law term, appears in the calculation of topological entanglement entropy. To make it explicit that the results above are TEE instead of the full entanglement entropy, alternatively we can use $Z_{1,0}(\tau)$ derived from supersymmetric localization method in Chern-Simons theory on 3-manifolds with boundaries \cite{3J}. Following the replica trick, we find exactly the same expression\footnote{The supersymmetric localization method involves boundary fermions. We need to remove the contribution from the boundary fermions to match with the partition function \eqref{eq:btz-partition}}. Since Chern-Simons theory is a topological quantum field theory, the resulting entanglement entropy is a TEE. The horizon area $r_+$ should be understood as a topological quantum number of the theory.

In the calculation of TEE between two disjoint thermal AdS$_3$'s, as stated in section \ref{sec:tads}, we have seen that a nonzero TEE is not enough to guarantee true nontrivial entanglement between two subregions because of the possible contribution from classical correlations. So we resort to the mutual information $I(A,B)$ between two single-sided black holes. We then need to find $S(A\cup B)$. Since in the Euclidean picture we are no longer at a pure state, it is not necessary that $S(A\cup B)$ vanishes, although $A\cup B$ consists the entire system.

\begin{figure}[htbp]
\centering
	\centering
	\begin{tikzpicture}[scale=0.7]
	\draw[thick,fill=lightgray] (0,0) circle [radius=2];
	\draw[thick,draw=blue,fill=gray] (0,0) circle[radius=1.5];
	\draw[thick,fill=white] (0,0) circle [radius=1];
	\draw[fill=white,thick] (-1.98,0.2)--(-1.7,0.2)--(-1.7,-0.2)--(-1.98,-0.2);
	\draw[thick] (8,0) arc (0:150:2 and 2)--(6,0);
	\draw[thick] (8,0) arc (0:-150:2 and 2)--(6,0);
	\draw[thick] (8,0) arc (0:-180:2 and 2);
	\draw[thick] (8.2,0) arc (0:-180:2.2 and 2.2);
	\draw[thick] (8,0)--(8.2,0);
	\draw[thick] (4,0)--(3.8,0);
	\draw[thick] (6,0)--(8,0);
	\node at (6,1) {$A$};
	\node at (6,-1) {$B$};
	\node at (6,-2.7) {$C$};
	\end{tikzpicture}
\caption{(Color online.) \textbf{Left:} Subregion $C$ is the small white square in the constant time slice. \textbf{Right:} One copy of $\rho_A$. The picture shows the disk perpendicular to the horizon. The thin layer surrounding the lower half circle corresponds $C$.}
\label{fig:cup}
\end{figure}
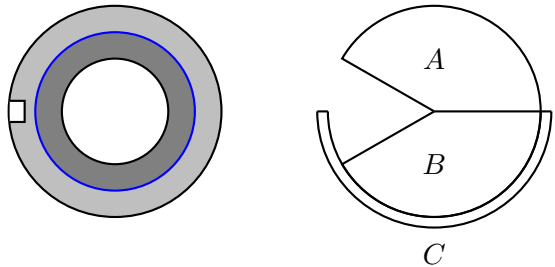

We start with bipartiting the system into $A\cup B$ and $C$ at $t=0$, as shown in Fig. \ref{fig:cup}. $C$ is a very small region whose area will finally be taken to zero. 

The glued manifold is a solid torus with modular parameter $2n\tau'$, exactly the same form with Fig. \ref{fig:beta-p}. The contributions from the $C$ vanish because they are still contractible in the glued manifold so we can safely take their area to be zero. Plugging \eqref{eq:btz-partition} into the replica trick formula \eqref{eq:replica}, we again obtain 
\begin{equation}
\label{eq:union}
    S_{BTZ}(A\cup B)=S_{BTZ}^{\text{thermal}}(A).
\end{equation}
So indeed the TEE of $A\cup B$ does not vanish. Combining these, we find that the mutual information is the same as the Bekenstein-Hawking entropy for a single-sided black hole:
\begin{equation}
    I(A,B)=S_{BTZ}(A)+S_{BTZ}(B)-S_{BTZ}(A\cup B)=S_{BTZ}^{\text{thermal}}(A).
\end{equation}

Note that, had we naively taken the full partition function of the eternal BTZ black hole to be $Z_{1,0}(\tau)^2$, namely, the two single-sided black holes are independent and non-entangled so that their partition functions can be multiplied together, then $S_{BTZ}(A\cup B)$ would have been twice $S_{BTZ}^{\text{thermal}}(A)$ and the mutual information would have vanished. So the nonzeroness of mutual information indicates nontrivial entanglement between $L$ and $R$.

There is still another surgery that can yield $S_{BTZ}^{\text{thermal}}(A)$: (1) restrict to the right single-sided black hole $R$ as the full spacetime, which is a solid torus with modular parameter $\tau$, obtained from rotating the constant time slice of it by $2\pi$; (2) thicken the horizon $S^1$ to a narrow annulus inside the spatial slice of the solid torus $R$; (3) calculate the TEE between the thin solid torus generated by thickened horizon, denoted by $B$, and the rest, denoted by $A$; (4) and finally take the limit that thickness of solid torus $B$ goes to zero.

\begin{figure}[htbp]
\centering
  \begin{tikzpicture}[scale=0.7]
  \draw[thick, fill=gray] (0,0) circle [radius=2];
  \draw[thick,fill=white] (0,0) circle [radius=1.2];
  \draw[thick, draw=blue] (0,0) circle [radius=1];
  \end{tikzpicture}
\caption{With the absence of black hole $L$, bipartitions of the constant time slices of black hole $R$ lead to $Z_{1,0}(n\tau)$ after gluing.
The gray area corresponds to subregion $A$, and the width of the annulus $B$ will be taken to zero.}
\label{fig:alternative}
\end{figure}
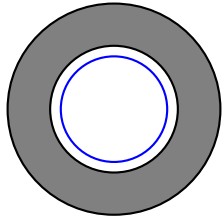

The bipartition of the constant slice in this case is sketched in Fig. \ref{fig:alternative}. In this bipartition, the obtained TEE is between the exterior and the interior of horizon, rather than that between two single-sided black holes. The glued manifold is again represented by $Z_{1,0}(n\tau)$ and the replica trick yields the Bekenstein-Hawking entropy.

We have thus come to a conclusion that the followings are equal:
\begin{itemize}
    \item[(a)] TEE between the two single-sided black holes,
    \item[(b)] TEE between the exterior and the interior of the horizon for a single-sided black hole,
    \item[(c)] thermal entropy of one single-sided black hole,
    \item[(d)] mutual information between the two single-sided black holes.
\end{itemize}
The equivalence of (a) and (c) supports the ER=EPR conjecture \cite{MaldacenaTFD,Raamsdonk,ER=EPR} in the Euclidean AdS$_3$ case.
The equivalence between (b) and (c) shows explicitly from the bulk perspective that one should view the thermal entropy of a black hole as entanglement entropy (see for example Ref. \cite{Solodukhin}). 

In general for a rotating BTZ black hole, although there is an inner horizon at $r=r_-$, the $z$-axis still represents the outer horizon at $r=r_+$ in the spherical coordinates \eqref{eq:spherical} for the upper $\mathbb{H}^3$. Hence, the replica trick described earlier still applies to a rotating BTZ black hole with modular parameter $\tau=\Phi+i\beta$, where $\Phi$ is the angular potential, the conjugate variable to angular momentum. Geometrically, we just need to put $r=|r_-|$ ``inside'' the inner edge of the constant time slice, so that it is not observable.\footnote{A similar situation will be described in appendix \ref{app:partition}.} 

\subsection{The Entangling-Thermal Relation}


In Ref. \cite{Azeyanagi}, the authors showed a relation \eqref{eq:thermal}
for a single-sided BTZ black hole between the entanglement entropy of CFT on the conformal boundary and the Bekenstein-Hawking entropy:
\begin{equation}
\label{eq:thermal}
\lim_{l\rightarrow 0}(S_{A}(L-l)-S_A(l))=S^{\text{thermal}},
\end{equation}
where $S_A(L-l)$ is the entanglement entropy of a subregion $A$ on the boundary 1+1d CFT with an interval length $(L-l)$, and $S^{\text{thermal}}$ is the thermal entropy in the bulk. In this section, we propose another similar but different Entangling-Thermal relation.

\begin{figure}[htbp]
	\centering
	\begin{tikzpicture}[scale=0.7]
	\draw[thick,fill=gray] (0,0) circle [radius=2];
	\draw[thick,blue,fill=white] (0,0) circle [radius=1];
	\filldraw[fill=white,thick] (-1.98,0.2)--(-1.7,0.2)--(-1.7,-0.2)--(-1.98,-0.2);
	\draw[thick,fill=gray] (6,0) circle [radius=2];
	\filldraw[fill=white,thick,draw=blue] (6,0) circle [radius=1];
	\filldraw[thick,fill=white] (4.02,0.2)--(3.7,0.2)--(3.7,-0.2)--(4.02,-0.2);
	\end{tikzpicture}
	\caption{Bipartition of the constant time slice. Left and right panels are equivalent.}
	\label{fig:ring-partition}
\end{figure}
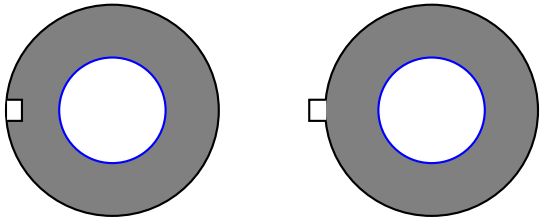

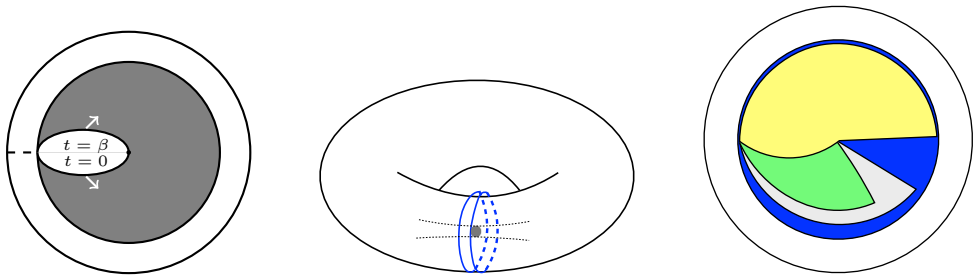
\begin{figure}[htbp]
\centering
\begin{tikzpicture}[scale=0.8]
	\filldraw[fill=gray, thick] (0,0) circle [radius=1.5];
	\draw[thick] (0,0) circle [radius=2];
	\filldraw[fill=white,thick] (0,0) arc (0:180:0.75 and 0.375);
	\filldraw[fill=white,thick] (0,0) arc (0:-180:0.75 and 0.375);
	\node at (-0.7, 0.1) {{\tiny $t=\beta$}};
	\node at (-0.7,-0.14) {{\tiny $t=0$}};
	\filldraw[fill=blue,thick] (0,0) circle [radius=0.02];
	\draw[dashed,thick] (-2,0)--(-1.5,0);
	\draw[->,thick,white] (-0.7,0.4)--(-0.5,0.6);
	\draw[->,thick,white] (-0.7,-0.4)--(-0.5,-0.6);
\end{tikzpicture}
\hskip 0.75 cm
\includegraphics[scale=0.3]{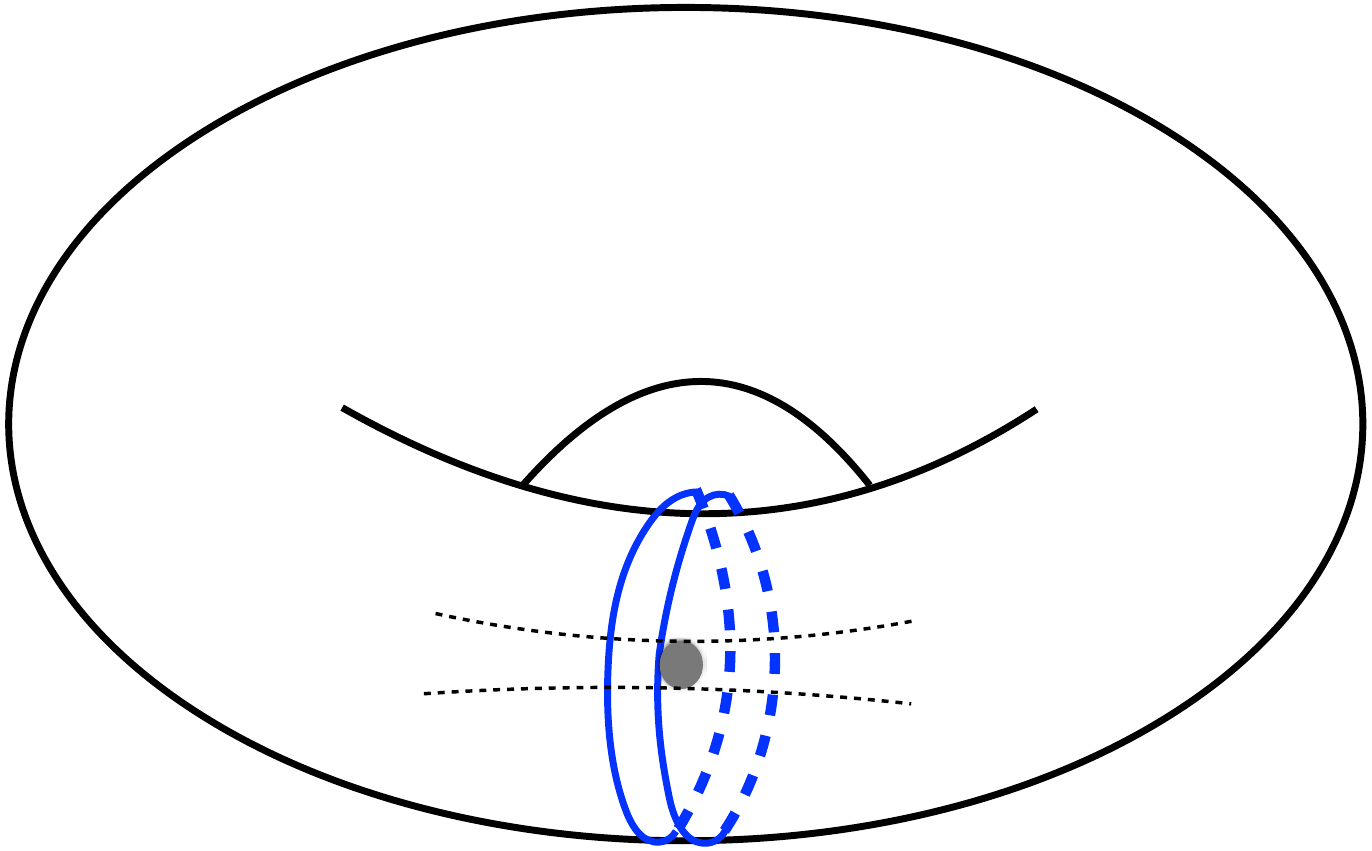}
\hskip 0.75 cm
\includegraphics[scale=0.25]{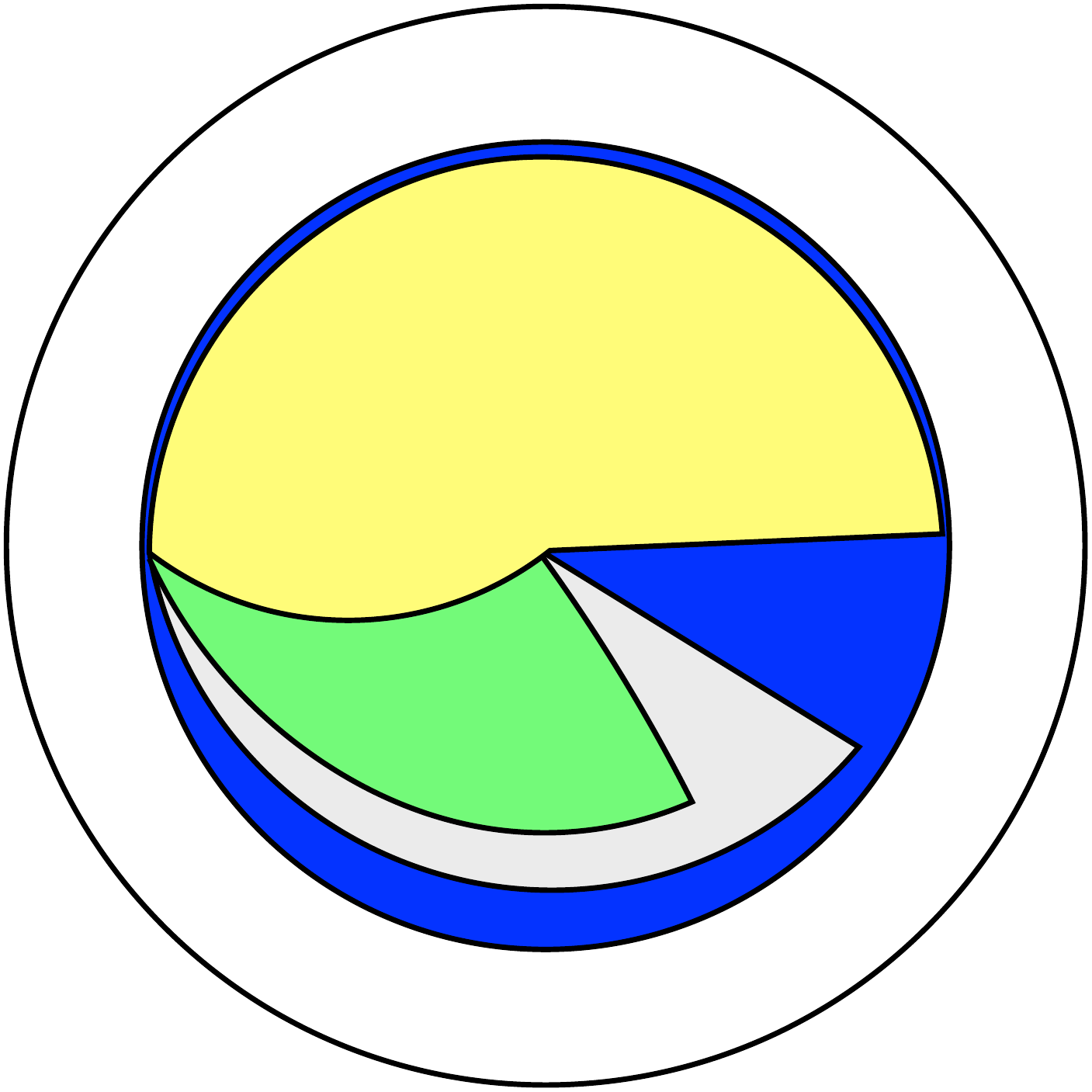}
\caption{(Color online) \textbf{Left:} The side view of $\text{tr}\rho_A$ for the ``ring''; the dashed line is only used to separate $t=0$ and $t=\beta$ ends of the grey region. \textbf{Middle:} the front view of $\text{tr}\rho_A$ for the ``ring'' configuration. \textbf{Right:} the side view of $\text{tr}\rho_A^4$ inside the ``ring'' of the first $\text{tr}\rho_A$.}
\label{fig:ring}
\end{figure}

We first consider the bipartition of the constant time slice as in Fig. \ref{fig:ring-partition} for a single-sided black hole. We put the separation between two subregions away from the horizon, so that region $B$ is the white contractible region in the left panel. The right panel is equivalent to the left one, and will be convenient for visualization of the gluing. We will call the glued manifold as the ``ring'', because after time evolution, region $B=\overline{A}$ will glue to itself and form a ring around the solid torus, shown in the middle panel of Fig. \ref{fig:ring}, where the small white part corresponds to the unglued part in its left panel. Hence, a single copy is the middle panel: away from the ring, the open wedge running around the longitude is the same as that in the left panel of Fig. \ref{fig:BTZsurgery}. 

Naively it seems that one is unable to glue $n$ copies of the above geometry, since the ring blocks a portion of the wedge's opening. However, there do exist a unique embedding from $n$ copies to $\mathbb{R}^3$ up to homotopy equivalence, as shown in the right panel of Fig. \ref{fig:ring}: one first stretches the grey region in the left panel to the blue area in right panel, and glue a second light grey copy so that its $t=0$ edge are glued to the $t=\beta$ edge of the blue copy; now one repeats this process for green and yellow regions and so on, still preserving the replica symmetry. Notice that rings from gray, green and yellow copies (color online) are not in this piece of paper, but on parallel planes above or below. Then one puts rings from each copy side by side on the boundary torus, which requires each ring to be infinitesimally thin since $n$ is arbitrarily large. The resultant manifold is again a solid torus of modular parameter $n\tau$. So the replica trick calculation follows the previous equation \eqref{eq:replica} and gives
\begin{equation}
\label{eq:ring}
\lim_{\text{Area}(\bar{A})\rightarrow 0} S(A)=S_{BTZ}^{\text{thermal}}.
\end{equation}

\begin{figure}[htbp]
	\centering
	\includegraphics[scale=0.2]{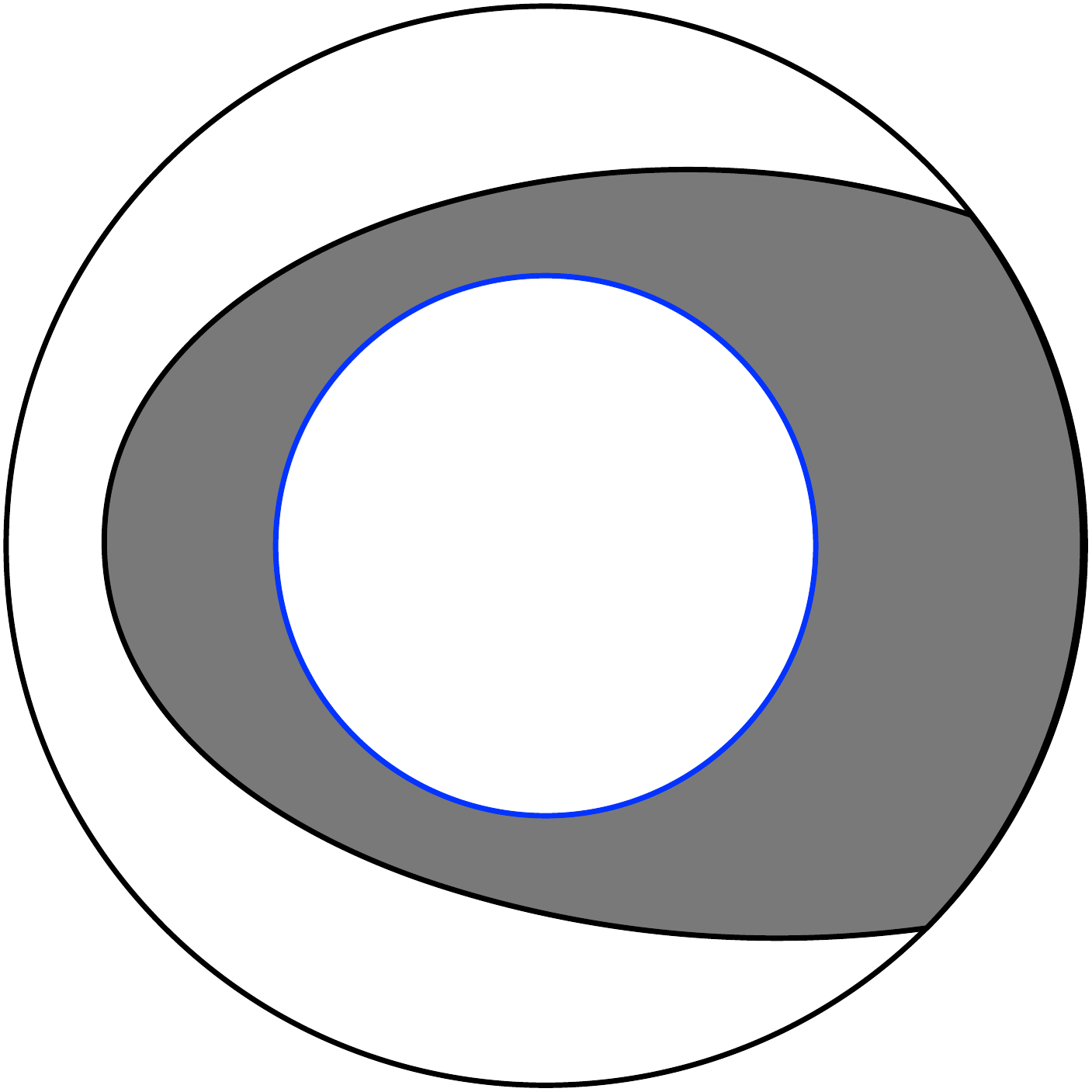}
	\caption{Another limit of the ring configuration.}
	\label{fig:other-limit}
\end{figure}

For completeness, we note that Fig. \ref{fig:ring} has another limiting case, where the width of the ring covers almost the entire longitudinal direction of the solid torus, and its depth occupies a considerable portion of the radial direction, as shown in Fig. \ref{fig:other-limit}. Now in order to put rings side by side upon gluing $n$ copies, we need to stretch the non-contractible direction for $n$ times to accommodate them, so that the resultant manifold is approximately a solid torus with modular parameter $\tau/n$. Now plug $Z_{1,0}(\tau/n)$ into \eqref{eq:replica}:
\begin{equation}
\begin{split}
\lim_{\text{Area}(A)\rightarrow 0}S(A) & =-\frac{d}{dn} \left( \frac{Z_{1,0}(\tau/n)Z_{1,0}(\tau)^{n}}{Z_{1,0}(\tau)^{2n}} \right)\Bigg|_{n=1}=\ln Z_{1,0}(\tau)+\tau\frac{d}{d\tau}Z_{1,0}(\tau).
\end{split}
\end{equation}

Using $Z_{1,0}(\tau)\approx e^{4k\pi/\beta}(1+2e^{-4\pi/\beta})$ again, we obtain
\begin{equation}
\label{eq:A}
\lim_{\text{Area}(A)\rightarrow 0}S(A)=2\left(\frac{4\pi}{\beta}+1\right)e^{-4\pi/\beta},
\end{equation}
which vanishes at high temperature. Note that here is no $k$-dependence, meaning we can observe the one-loop effect directly.

\begin{figure}[htbp]
\centering
\label{fig:complement}
\begin{tikzpicture}[scale=0.7]
	\filldraw[thick,fill=white] (6,0) circle [radius=2];
	\filldraw[fill=white,thick,draw=blue] (6,0) circle [radius=1];
	\draw[thick,fill=gray] (4.02,0.2)--(3.7,0.2)--(3.7,-0.2)--(4.02,-0.2);
\end{tikzpicture}
\hskip 1cm
\includegraphics[scale=0.2]{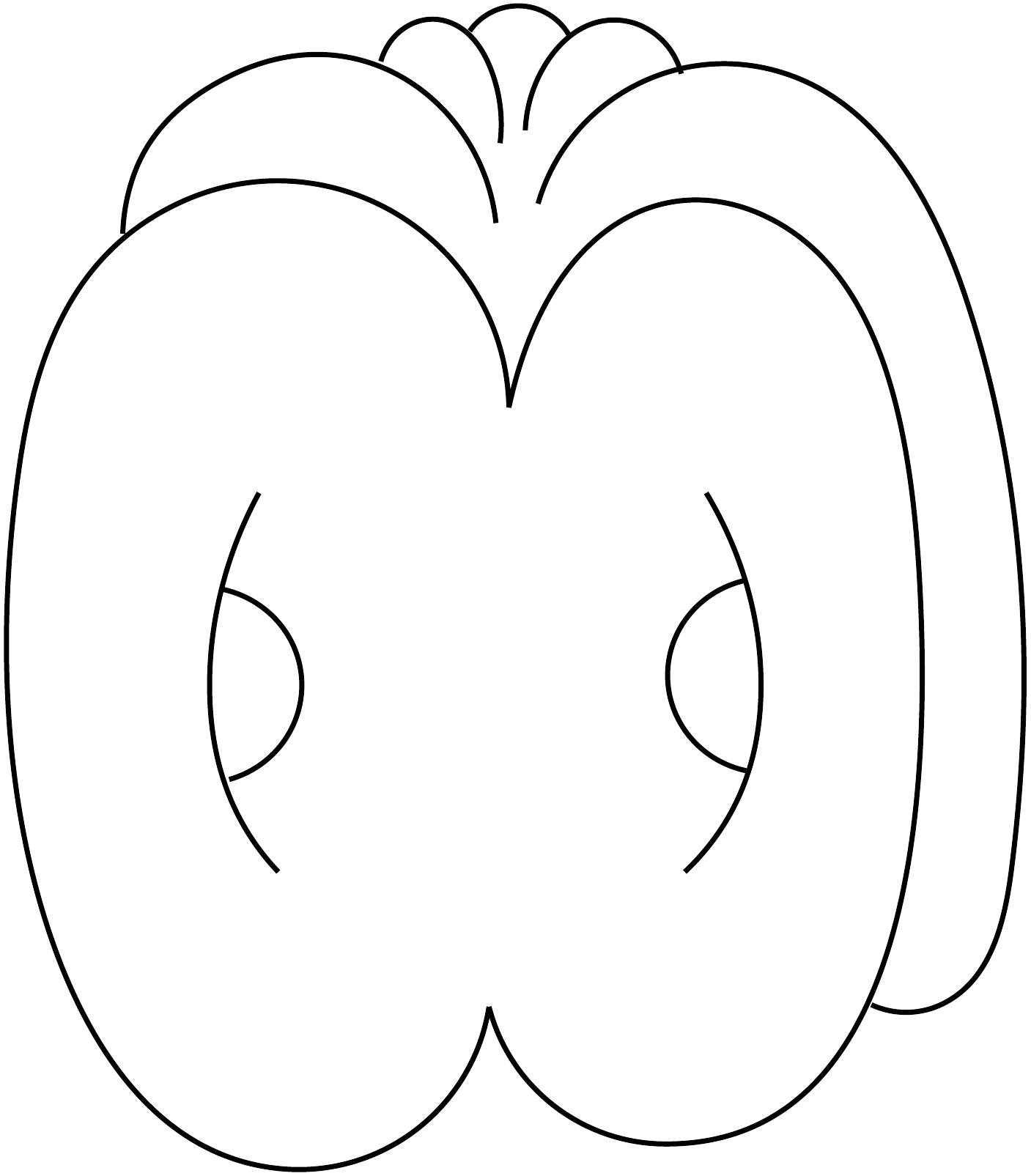}
\caption{\textbf{Left:} The complementary bipartition which leads to $S_{\bar{A}}$. \textbf{Right:} The glued manifold is a vertical bouquet-like handlebody.}
\label{fig:vertical}
\end{figure}
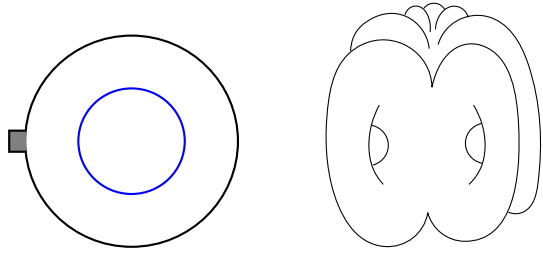

Now we consider the complementary bipartition to Fig. \ref{fig:ring}, shown in Fig. \ref{fig:vertical}, where the grey region is $\bar{A}$ in Fig. \ref{fig:ring}. The gluing here is simple: since the unglued cut in the grey region $A$ is parallel to the longitude, $n$ copies should be arranged around a virtual axis tangent to the annulus. The resultant manifold is a vertical $n$-handlebody. 

One can calculate the corresponding TEE following a parallel procedure in the calculation of thermal \ads in section \ref{sec:tads}. The partition function of the glued manifold is
\begin{equation}
Z(n)= \prod_{\gamma~\text{prim.}}\prod_{m=1}^\infty |1-q_\gamma^m|^{24k}\times \prod_{\gamma~\text{prim.}} \prod_{m=2}^\infty \frac{1}{|1-q_\gamma^m|},
\end{equation}
where the first and second factors come from tree level and one-loop, respectively. The products are over primitive conjugacy classes of $\gamma\in\Gamma$. In the high-temperature regime, this expression can be simplified by the single-letter word approximation $\prod\limits_{\gamma~\text{prim.}} |1-q_\gamma| \approx |1-q_1'|^{2n}$, so that
\begin{equation}
Z(n,q_1')\approx \frac{|1-q_1'|^{48nk}}{|1-q_1'^2|^{2n}}.
\end{equation}
Here $q_1'$ can be obtained from $q_1$ in \eqref{eq:tads-q1} using a modular transformation,
\begin{equation}
\label{eq:tads-q2}
q_1'(n,a)=\frac{\sinh^2(\pi a/\beta)}{n^2\sinh^2(\pi a/n\beta)}e^{-2\pi/\beta}.
\end{equation}

The replica trick then gives
\begin{equation}
S(\bar{A})=-\frac{d}{dn}\left[\frac{Z(n,q_1'(n))}{Z(1,q_1'(1))^n}\right]\Bigg|_{n=1}.
\end{equation}
This is explicitly written as
\begin{equation}
\label{eq:entangling}
S(\bar{A})=96k\left(\frac{\pi a}{\beta}-2\right)e^{-2\pi/\beta}+8(12k-1)\left(\frac{\pi a}{\beta}-2\right)e^{-4\pi/\beta}+O(e^{-6\pi/\beta}).
\end{equation}
We now take the limit $a\rightarrow 0$ because this corresponds to the limit where the grey region in Fig. \ref{fig:vertical} goes to zero, so that:
\begin{equation}
\label{eq:Abar}
\lim_{\text{Area}(\bar{A})\rightarrow 0}S(\bar{A})\equiv\lim_{a\rightarrow 0}S(\bar{A})=-192ke^{-2\pi/\beta}-16(12k-1)e^{-4\pi/\beta}+O(e^{-6\pi/\beta}),
\end{equation}
which vanishes at high temperature. The infinitesimally negative value is a quirk due to approximation on $q_{\gamma}$'s.

Combining equations \eqref{eq:ring} and \eqref{eq:Abar}, one obtains the {\it Entangling-Thermal relation:}
\begin{equation}
\label{eq:entangling-thermal}
\lim_{\text{Area}(\bar{A})\rightarrow 0}[S(A)-S(\bar{A})]=S_{BTZ}^{\text{thermal}},
\end{equation}
We give this relation a different name from the two-dimensional thermal entropy relation in the dual CFT calculation \eqref{eq:thermal} because this is not merely a generalization of it in one higher dimension. The thermal entropy relation \eqref{eq:thermal} relates the entanglement entropy on the dual CFT with the thermal entropy of black hole in the bulk, while the entangling-thermal relation connects the topological entanglement entropy and thermal entropy both in the bulk gravitational theory. Additionally, the explanation for thermal entropy relation relies on the geometrical detail (minimal surfaces) in the bulk \cite{Azeyanagi}, while the entangling-thermal relation is of topological origin. In the first bipartition in Fig. \ref{fig:ring}, subregion $A$ sees the non-contractible loop and the nontrivial flux threading through the hole inside the annulus. In the second bipartition in Fig. \ref{fig:vertical}, subregion $A$ does not completely surround the non-contractible circle, i.e. the horizon. The difference between them this characterizes the non-contractible loop.

Finally we remark that there are several cases in which gluing procedures are not available. The no-gluing criterion being that, as long as the boundary of a subregion is contractible and not anchored on the boundary $S^1$, the spatial slice is not $n$-glueable. Also, a single copy in which glued region $B$ completely surrounds region except for the inner edge is not $n$-glueable.



\section{Summation over Geometries}
\label{sec:whole}

The partition functions of thermal \ads $Z_{0,1}(\tau)$ and BTZ black hole $Z_{1,0}(\tau)$ are not modular-invariant by themselves. To obtain the full modular-invariant partition function, one needs to sum over the pair of parameters $(c,d)$ for $Z_{c,d}$. This can alternatively be written as the summation over modular transformations of $Z_{0,1}$ as follows:
\begin{equation}
\label{eq:sum}
Z(\tau)=\sum_{\Gamma_{\infty}\textbackslash SL(2,\mathbb{Z})}Z_{c,d}(\tau)=\sum_{\Gamma_{\infty}\textbackslash SL(2,\mathbb{Z})}Z_{0,1}\left(\frac{a\tau+b}{c\tau+d}\right).
\end{equation}
where $\Gamma_{\infty}\textbackslash SL(2,\mathbb{Z})$ denotes the left coset of $SL(2,\mathbb{Z})$ by $\Gamma_{\infty}$ \cite{Manschot}, the translational subgroup generated by $2\times2$ matrices $\left(\begin{matrix}
1 & r\\
0 & 1
\end{matrix}\right)$ with action $\tau\rightarrow\tau+r$. Solid torus filling and Schottky parametrization are invariant under $\Gamma_{\infty}$, and the summation over coset is to make the full partition function invariant under both $T:\tau\rightarrow \tau+1$ and $S:\tau\rightarrow-1/\tau$.

Note that in the previous sections we have used $Z_{c,d}(\tau)=Z_{c,d}(\tau,\bar{\tau})$ as the shorthand for the product of holomorphic and anti-holomorphic pieces, whereas in this section we return to the notation that $Z_{c,d}(\tau)$ describes the holomorphic part of the partition function only. The anti-holomorphic part can easily be found as $\bar{Z}(\bar{\tau})$ and $Z(\tau,\bar{\tau})=Z(\tau)\bar{Z}(\bar{\tau})$.

Modular-invariant partition function of the form \eqref{eq:sum} is unique for the most negative cosmological constant $(k=1)$ \cite{Hohn,Witten0706} and was investigated in more general situations $(k>1)$ in \cite{MaloneyWitten}. An important theorem due to \cite{Hohn} is that the moduli space of Riemann surfaces of genus one is itself a Riemann surface of genus zero, parametrized by the $j$-function. Consequently, any modular-invariant function can be written as a function of it. The $J$-function is defined as 
\begin{equation}
\begin{split}
J(\tau)&\equiv\frac{1728g_2(\tau)^3}{g_2(\tau)^3-27g_3(\tau)^2}-744\\
&=q^{-1}+196884q+21493760q^2+864299970q^3+20245856256q^4+\dots
\end{split}
\end{equation}
where $q=e^{2\pi i\tau}$ as usual, and $g_2(\tau)\equiv60G_4(\tau)$ and $g_3(\tau)\equiv140G_6(\tau)$, where $G_{2k}$ are holomorphic Eisenstein series of weight $2k,\,k\geq2$, defined as $G_{2k}\equiv\sum_{(m,n)\neq (0,0)}(m+n\tau)^{-2k}.$

Since the pole in the full partition function $Z(q)$ at $q=0$ is of order $k$ (due to the holomorphic tree-level contribution of thermal AdS$_3$, $q^{-k}$), it must be a polynomial in $J$ of degree k,
\begin{equation}
Z(q)=\sum_{j=0}^k a_i J^i=\sum_n c(k,n)q^n.
\end{equation}
For $k=1$ we simply have $Z(q)=J(q)$.
The coefficients of $J(q)$ in front of $q^n$ was known to be intimately related to the dimensions of irreducible representations of the monster group $\mathbb{M}$, the largest sporadic group. It has $2^{46}\cdot3^{20}\cdot5^9\cdot7^6\cdot11^2\cdot13^3\cdot17\cdot19\cdot23\cdot29\cdot31\cdot41\cdot47\cdot59\cdot71\approx8\times10^{53}$ group elements and $194$ conjugacy classes. Dimensions of the irreducible representations of the monster group can be found in the first column of its character table \cite{Atlas}: 1, 196883, 21296876, 842609326, 18538750076, 19360062527 $\dots$. 

After John McKay's observation $196884=1+196883$, Thompson further noticed \cite{Thompson}:
\begin{equation}
\label{eq:Thompson}
\begin{split}
& 21493760 = 1 + 196883 + 21296876,\\
& 864299970 = 2\times1 + 2\times196883 + 21296876 + 842609326,\\
& 20245856256 = 2\times1 + 3\times196883 + 2\times21296876 + 842609326 + 19360062527.
\end{split}
\end{equation}
This phenomenon is dubbed ``monstrous moonshine'' by Conway and Norton \cite{ConwayNorton}, later proved by Borcherds \cite{Borcherds}. 

Ref. \cite{Witten0706} conjectures that for cosmological constant $k\equiv l/16G\in \mathbb{Z}$, quantum 3d Euclidean pure gravity including BTZ black holes can be completely described by a rational CFT (RCFT) called extremal self-dual CFT (ECFT) with central charge $(c_L,c_R)=(24k,24k)$, which is factorized into holomorphic and an anti-holomorphic pieces. An ECFT is a CFT whose lowest dimension of primary field is $k+1$, and it has a sparsest possible spectrum consistent with modular invariance, presenting a finite mass gap. The only known example is the $k=1$ one with a monster symmetry, constructed by Frenkel-Lepowsky-Meurman (FLM) \cite{FLM} to have partition function as $J(q)$, but its uniqueness has not been proved. The existence of ECFTs with $k>1$ is conjectured to be true \cite{Witten0706} and is also an active open question \cite{bootstrap,Gaiotto}. 

In this section we will mainly focus on the $k=1$ case.

\subsection{TEE for the Full Partition Function}

The modular-invariant partition function is still defined on a solid torus. We will again consider the bipartition that separate the two single-sided black holes, similar to the section \ref{subsec:btz-erepr}. It is justified in appendix \ref{app:partition} that one can still cut $SL(2,\mathbb{Z})$ family of BTZ black holes along their outer horizons, which lie in the core of the solid torus. So one just needs to plug the partition function $J(q)$ into the replica trick formula. 

At low temperatures, $q=e^{-2\pi\beta}$ is small, so that the full partition function will be dominated by the $q^{-1}$ term with almost trivial thermal entropy and TEE, trivial in the sense that there are no tree-level contributions. At high temperatures, richer physics is allowed. 
Below we calculate the TEE of the full partition function in this regime.

Generally, the coefficient in front of $q^n$ in the partition function $Z(q)$ for any $k$ can be written as 
\begin{equation}
\label{eq:c}
c(k,n)=\sum_{i=0}^{193}\textbf{m}_i(-k,n)d_i,
\end{equation}
where each $d_i$ is the dimension of the corresponding irreducible representations $M_i$ of $\mathbb{M}$, and $\textbf{m}_i(-k,n)$ is the multiplicity of the irreducible representation $M_i$ in the decomposition similar to \eqref{eq:Thompson}. It is guaranteed to be a non-negative integer. At large $n$, $\textbf{m}_i(-k,n)$ has the following asymptotic form \cite{Duncan1},
\begin{equation}
\label{eq:asymptotic}
\textbf{m}_i(-k,n)\sim\frac{d_i|k|^{1/4}}{\sqrt{2}|\mathbb{M}||n|^{3/4}}e^{4\pi\sqrt{|kn|}}.
\end{equation}

Now we restrict to the $k=1$ case and let $n$ to be a variable. 
Taking care of the anti-holomorphic part, the replica trick \eqref{eq:replica} gives the following TEE, which is again equal to the thermal entropy \begin{equation}
\label{eq:full}
S_{\text{full}}(A)=S_{\text{full}}^{\text{thermal}}=2\ln J(q)-2\beta J(q)^{-1}\frac{\partial J(q)}{\partial\beta}.
\end{equation}
Note that this is again the same as the expression for calculation of thermal entropy in the canonical ensemble. (Using $\beta=l/r_+=1/\sqrt{M}= 1/\sqrt{n}$, n is viewed as a function of $\beta$ so the second term in \eqref{eq:full} is nonzero.)
The computation of $S_{A\cup B}$ for the entire $SL(2,\mathbb{Z})$ family of black holes is also similar to that of $M_{1,0}$ calculated in section \ref{subsec:btz-erepr}. The result is again equal to the thermal entropy, based on the fact that the $SL(2,\mathbb{Z})$ family of black holes are all solid tori with horizons living in the core. This implies that the system is again in a mixed state due to the Euclideanization, as expected in \cite{Page,Hawking}. The mutual information $I(A,B)$ is also the thermal entropy, parallel to the discussion in section \ref{sec:btz}.

In the high-temperature expansion, we only take the $q^n$ term $J_n(q)$ from the summation in $J(q)$ to calculate TEE because the desired term has a coefficient exponentially larger than those at lower temperatures\footnote{We will take into account all terms of $J(q)$ in appendix \ref{app:J}.}:
\begin{equation}
\label{eq:J}
J_n(q) = \sum_{i=0}^{193}\frac{d_i^2}{|\mathbb{M}|}\frac{e^{4\pi\sqrt{n}}}{\sqrt{2}n^{3/4}}q^n.
\end{equation}
Mathematically the two copies of $d_i$ in $d_i^2$ are both the quantum dimension of irreducible module $M_i$ of the monster group, which will be explained in detail later in section \ref{sec:qdim}. But physically they have different origins: one is the contribution from a single $M_i$ as shown in equation \eqref{eq:c}, while the other is probability amplitude for $M_i$ to appear in the summation as in equation \eqref{eq:asymptotic}. Namely, there is a correspondence between the partition function $J(q)$ and a pure state in the bulk, which is a superposition of different $M_i$'s:
\begin{equation}
\label{eq:Baez}
|\Psi\rangle = \sum_{i=0}^{193} \frac{d_i}{\sqrt{|\mathbb{M}|}} |i, i^*\rangle.
\end{equation}
In analogy to topological phases, the state is a {\it maximally-entangled state of $194$ types of anyons} labelled by the irreducible representations of the Monster group $\mathbb{M}$. The $d_i$ that appears explicitly in \eqref{eq:Baez} corresponds to that in \eqref{eq:asymptotic}, whereas $|i, i^*\rangle$ means a quasiparticle-antiquasiparticle pair labeled by $M_i$ and contributes another $d_i$, which correspond to the one in \eqref{eq:c}. In Ref. \cite{Baez}, the authors proposed from abstract category theory, that the ER=EPR realization in the context of TQFT should be exactly of the form \eqref{eq:Baez}. We will show later that this specific maximally-entangled superposition is the bulk TQFT version of the thermofield double state on the dual CFTs.

Applying to equation \eqref{eq:J} the identity for finite groups: $\sum_i d_i^2=|\mathbb{M}|$, we arrive at
\begin{equation}
J_n(q)=\frac{e^{4\pi\sqrt{n}}}{\sqrt{2}n^{3/4}}q^n=\frac{1}{\sqrt{2}}\beta^{3/2}e^{2\pi/\beta}.
\end{equation}
Plugging into \eqref{eq:full} and taking into account the anti-holomorphic part, we again recover the Bekenstein-Hawking entropy:
\begin{equation}
\label{eq:J-S}
S_{\text{full}}(A)
=\frac{8\pi}{\beta}+3\ln\beta-\ln2-3.
\end{equation}
The first three terms agree with Witten's asymptotic formula for Bekenstein-Hawking entropy \cite{Witten0706}, and provides an additional term $-3$. Remarkably, the ``anyons'' become invisible in TEE after the summation over $i$. This is exactly due to the appearance of the maximally-entangled superposition in equation \eqref{eq:Baez}. Had we taken another state where only one single $M_j$ appears with probability amplitude $1$ and all the others appear with amplitude $0$, then the corresponding term would have been proportional to $\ln \left(d_j/\sqrt{|\mathbb{M}|}\right)$. The latter matches with the entanglement entropy calculations in Refs. \cite{He1,Caputa,He2} for an excited state labeled by $j$ in a rational CFT.\footnote{This disappearance of ``anyons'' in the TEE for a maximally-entangled superposition is also expected in the context of topological phases, see equation (40) of Ref. \cite{Fradkin}, where one takes $|\psi_j|$ there to be $d_j/D$. }

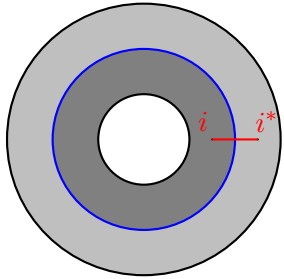
\begin{figure}[htbp]
\centering
\begin{tikzpicture}[scale=0.6]
	\filldraw[fill=lightgray, draw=black, thick] (0,0) circle [radius=3];
	\filldraw[fill=gray,  draw=blue, thick] (0,0) circle [radius=2];
	\filldraw[fill=white,  thick] (0,0) circle [radius=1];
	\filldraw[fill=red] (1.5,0) circle [radius=0.01];
	\filldraw[fill=red] (2.5,0) circle [radius=0.01];
	\draw[thick,red] (1.5,0)--(2.5,0);
	\node at (1.3,0.4) {\zhuxi{$i$}};
	\node at (2.7,0.4) {\zhuxi{$i^*$}};
	\end{tikzpicture}
	\caption{(Color online) Constant time slice of the eternal BTZ black hole as in Fig.\ref{fig:ConstantBTZ}. The Wilson line corresponding to the quasiparticle-antiquasiparticle pair $i$, $i^*$ intersects with horizon both on the constant time slice and in the $3$d bulk.}
	\label{fig:Wilson}
\end{figure}

In our case, the creation of the quasiparticle-antiquasiparticle pair $i$ and $i^*$ can be represented by a Wilson line, as shown in Fig. \ref{fig:Wilson}. The Wilson line intersects the non-contractible loop of the solid torus, i.e. the horizon, which is the reason why it can be detected by a cut along the horizon.

To make full understanding of the ``anyon'' picture, we rewrite state \eqref{eq:Baez} as
\begin{equation}
\label{eq:moonshine-double}
|\Psi\rangle=\frac{1}{\sqrt{J(q)}}\sum_{i=0}^{193}e^{-\frac{\beta}{2}E_i}|i,i^*\rangle, 
\end{equation}
where the energy level corresponding to the anyon pair $i, i^*$ is described by the quantum dimension of $M_i$:
\begin{equation}
\label{eq:energy}
E_i=-\frac{1}{\beta}\ln\left[\frac{d_i^2}{|\mathbb{M}|}J_n(q)\right].
\end{equation}
Denoting $|i,i^*\rangle\equiv|i\rangle|i^*\rangle$, one can trace over all the $|i^*\rangle$'s and obtain the reduced density matrix 
\begin{equation}
    \rho_A=\sum_i e^{-\beta E_i}|i\rangle\langle i|,
\end{equation}
which is just the thermal density matrix for anyons, and different types of anyons $i$ form an ensemble. Using the expression for energy levels \eqref{eq:energy}, the entanglement entropy between the anyon pair can be easily calculated as
\begin{equation}
    S_\Psi(A)=S^{\text{thermal}}(A)=S_{\text{full}}(A),
\end{equation}
where we have added the anti-holomorphic contribution. Thus the state \eqref{eq:moonshine-double} has the similar property as the thermofield double state in that the entanglement entropy between the quasiparticle-antiquasiparticle pair is equal to the thermal entropy of one quasiparticle. We call this state in the 3d bulk as the \emph{Moonshine double state}, in which the pair of anyons are separated by the horizon, just like the two single-sided black holes $L$ and $R$ are separated by it.

Unfortunately it has a shortcoming: as a pure state, the Moonshine double state above cannot reproduce the result of nonzero $S(A\cup B)$ \eqref{eq:union}.
To account for this, one could modify the final total quantum state as
\begin{equation}
\rho=|\tilde{\Psi}\rangle\langle\tilde{\Psi}|\otimes\rho_{\text{th}},
\end{equation}
where the modified moonshine double state now reads $|\tilde{\Psi}\rangle=\frac{1}{\sqrt[4]{J(q)}}\sum_{i=0}^{193}e^{-\frac{\beta}{2}\tilde{E}_i}|i,i^*\rangle$ with $\tilde{E}_i=-\frac{1}{\beta}\ln\left[\frac{d_i^2}{|\mathbb{M}|}J_n(q)^{1/2}\right]$. These energy level lead to the partition function $Z(q)=J(q)^{1/2}$. When one bipartites the system into two two single-sided black holes $A$ and $B$, one can see from straightforward computation that $|\tilde{\Psi}\rangle$ will contribute half of Bekenstein-Hawking entropy. The newly introduced $
\rho_{\text{th}}$ is purely thermal and exhibits no non-local correlations between $A$ and $B$, so that its von Neumann entropy is extensive and scales with volume. When one bipartites the system into the two single-sided black holes $A$ and $B$, it will give half of the Bekenstein-Hawking entropy. Combining the contribution from $|\tilde{\Psi}\rangle$, we recover $S_{\tilde{\Psi}}(A)=S^{\text{thermal}}(A)$, the Bekenstein-Hawking entropy. When considering $S(A\cup B)$, the modified moonshine double state contributes nothing as a pure state, while the result for $\rho_{\text{th}}$ is simply $S^{\text{thermal}}(A)$, matching with the calculations in \eqref{eq:union}.

Another caveat is that since $\ln J$ is approximately the Bekenstein-Hawking entropy, and the leading term in $E_i$ scales with $-\beta^{-2}\sim-n$. So in order to have a genuine quantum theory, our theory has to have a UV cutoff scale at a certain $n$.  

Apart from the asymptotic expression \eqref{eq:asymptotic} which gives rise to the tree-level Bekenstein-Hawking entropy, there is the remainder formula \cite{Remainder} for coefficients of $q^n$ in the whole partition function $J(q)$  which is possibly related to the one-loop contribution to TEE. For general $k\in\mathbb{Z}_+$, the remainder formula reads
\begin{equation}
\begin{split}
\label{eq:remainder}
c(k,n)&=\frac{ke^{4\pi\sqrt{kn}}}{\sqrt{2}(kn)^{3/4}}\left[1+\sum_{m=1}^{p-1}\frac{(-1)^m(1,m)}{(8\pi\sqrt{kn})^m}+\frac{r_p(kn)}{(kn)^{p/2}}+\frac{\sqrt{2}n^{3/4}}{e^{4\pi\sqrt{n}}}S(k,n)\right.\\
&\left.+\frac{1}{k^{1/4}}\sum_{1\leq r<k}\frac{r^{1/4}a_{-r}(k)}{e^{4\pi\sqrt{n}(\sqrt{k}-\sqrt{r})}}\left(1+\sum_{m=1}^{p-1}\frac{(-1)^m(1,m)}{(8\pi \sqrt{kn})^m}+\frac{r_p(kn)}{(kn)^{p/2}}+\frac{\sqrt{2}n^{3/4}}{e^{4\pi\sqrt{n}}}S(k,n)\right)\right],
\end{split}
\end{equation}
where $p(x)$ is the integer partition of $x\in\mathbb{Z_+}$, and
\begin{equation}
    \begin{split}
        & (1,k)\equiv\prod_{j=0}^{k-1}\frac{4-(2j+1)^2}{4^kk!},\quad
a_r(k)\equiv p(r+k)-p(r+k-1),\\
        & |r_p(n)|\leq\frac{|(1,p)|}{\sqrt{2}(4\pi)^p}+62\sqrt{2}e^{-2\pi\sqrt{n}}n^{p/2},\quad
0<\frac{\sqrt{2}n^{3/4}}{e^{4\pi\sqrt{n}}}S(k,n)\leq \frac{1}{4}\zeta^2\left(\frac{3}{2}\right)\frac{(rn)^{3/2}}{e^{4\pi\sqrt{rn}}}.
    \end{split}
\end{equation}
To check this claim, one could restrict to the $k=1$ monstrous case and plug this expression into \eqref{eq:full}. 
Alternatively one may fix $n$ and view the $c(k,n)$ as number of possible microstates at fixed energy, i.e. in the micro-canonical ensemble. One then performs a unilateral forward Laplace transform to return to canonical ensemble and then plug it to \eqref{eq:full}. Computations in both methods are in general complicated, and we do not pursue it here.

We provide another perspective towards the loop contribution in appendix \ref{app:J} by plugging in the whole $J$ function instead of only one large $n$ term. We observe that the loop correction is negative, consistent with both the thermal section \ads \ref{sec:tads} and the BTZ case in section \ref{sec:btz}.

\subsection{$d_i$ as Quantum Dimensions}
\label{sec:qdim}

In this section we provide more mathematical details and show that $d_i$ is the quantum dimension of the irreducible module $M_i$ of $|\mathbb{M}|$. An ECFT at $k=1$ is a special vertex operator algebra (VOA) $V^\natural$ whose automorphism group is the Monster group $\mathbb{M}$. This VOA, also known as the moonshine module \cite{FLM}, is an infinite-dimensional graded representation of $\mathbb{M}$ with an explicit grading:
\begin{equation}
V^\natural=\bigoplus_{n=-1}^{\infty} V_n^\natural,
\end{equation}
where every $V_n^\natural$ is an $\mathbb{M}$-module, called a homogeneous subspace. It can be further decomposed into
\begin{equation}
\label{eq:decomposition}
V_n^\natural\simeq \bigoplus_{i=0}^{193} M_i^{~\oplus\mathbf{m}_i(-1,n)},
\end{equation}
with $M_i$ labeling the irreducible $\mathbb{M}$-modules, and $\mathbf{m}_i(-1,n)$ is the multiplicity of $M_i$. This is the same multiplicity that appears in \eqref{eq:c}. (For ECFTs with general $k$, we have a tower of moonshine modules \cite{Duncan1} $V^{(-k)}=\bigoplus_{n=-k}^{\infty}V_n^{(-k)}$, where $V^{(-k)}_n$'s are all irreducible $\mathbb{M}$-modules. For each summand, one can similarly define $\textbf{m}_i(-k,n)$ as the multiplicity of the $\mathbb{M}$-modules $M_i$ in $V_n^{(-k)}$, so that $V_n^{(-k)}\simeq\bigoplus_{i=0}^{193}M_i^{\oplus\textbf{m}_i(-k,n)}.$)

Since we restrict to the holomorphic part of $Z(\tau,\overline{\tau})$ in this section, the entire dual CFT contains the ECFT above as a holomorphic piece. Furthermore, it is diagonal, i.e. its Hilbert space is a graded sum of tensor products of holomorphic and anti-holomorphic sectors: 
\begin{equation}
\mathcal{H}\cong\bigoplus_{\alpha\in\mathbb{C}}\mathcal{M}_{\alpha}\otimes\overline{\mathcal{M}}_{\alpha},
\end{equation}
where $\mathcal{M}_{\alpha}$ and $\overline{\mathcal{M}}_{\alpha}$ are indecomposable representations of right and left Virasoro algebras. Since Virasoro action is built into the VOA axioms \cite{axiom}, these are also modules of the right and left monstrous VOAs, so $V^{\natural}$ admits induced representations from representations of the Virasoro algebra \cite{Ben-Zvi}. Obviously there are infinite number of Virasoro primaries, and $V^{\natural}$ is not an RCFT in this sense. However, $V^{\natural}$ is a typical example of a holomorphic/self-dual VOA, i.e. there is only one single irreducible $V^{\natural}$-module which is itself. Knowing that there is only one VOA-primary, one can reorganize Virasoro fields in $\mathcal{M}_\alpha$ and $\overline{\mathcal{M}}_\alpha$ into irreducible representations of $V^\natural$, by introducing the graded dimension of the $V^\natural$-module $N$, defined as
\begin{equation}
\text{ch}_q N\equiv \text{tr}_Nq^{L_0}=\sum_{n=0}^{\infty}\dim N_nq^n,
\end{equation}
where $L_0$ is the usual Virasoro generator and $N_n$'s are homogeneous subspaces of $N$ labelled by eigenvalues of $L_0$. (Note that we have omitted the overall prefactor $q^{-c/24}$ often appeared in literature.) The above procedure is similar to regourpong infinite Virasoro primaries in WZW models into finite Kac-Moody primaries.

To explain the $d_i$ appearing in \eqref{eq:J}, it is natural to consider quantum dimensions associated to $V^{\mathbb{M}}$ consisted of fixed points of the action by $\mathbb{M}$ on $V^{\natural}$. By theorem 6.1 in \cite{Dong}, we have the following decomposition of $V^\natural$
\begin{equation}
V^{\natural}\simeq\bigoplus^{194}_{i=1}V^{\mathbb{M}}_i\otimes M_i
\end{equation}
as $V^{\mathbb{M}}\times\mathbb{M}$-modules, for the 194 $V^{\mathbb{M}}$-submodules $V^{\mathbb{M}}_i$ in $V^{\natural}$ with $V^{\mathbb{M}}=V^{\mathbb{M}}_1$, where $M_i$ denotes an irreducible module for $\mathbb{M}$ with character $d_i$. This $V^{\mathbb{M}}$ is a sub-VOA of $V^{\natural}$ of CFT type \cite{Gannon}, and is called the \textit{monster orbifold}, because it is obtained from orbifolding $V^{\natural}$ by its automorphism group $\mathbb{M}$ \cite{Norton}, in the same sense as orbifolding the Leech lattice VOA by $\mathbb{Z}/2\mathbb{Z}$ in the FLM construction. 

The standard definition of the quantum dimension of a VOA-module $N$ with respect to a general VOA $V$ is \cite{Dong}
\begin{equation}
\label{eq:character}
\text{qdim}_VN=\lim_{q\rightarrow1^-}\frac{\text{ch}_qN}{\text{ch}_qV}.
\end{equation}
The quantum dimensions of submodules of orbifold VOA $V^G$ obtained from orbifolding $V$ by a subgroup $G\subseteq\text{Aut}(V)$ only recently found their applications in quantum Galois theory \cite{Dong}. In our case, the quantum dimensions of all $V^{\mathbb{M}}_i$'s with respect to $V^{\mathbb{M}}$ were first calculated to be $\text{qdim}_{V^{\mathbb{M}}}V^{\mathbb{M}}_i=d_i$ in \cite{Duncan1}, using the asymptotic formula for multiplicities of $\mathbb{M}$-module $M_i$ in Fourier coefficient of $j$-invariant, bypassing the knowledge of $V^{\mathbb{M}}$'s rationality, which is still only conjectured to be true. 

The remaining question is to define in parallel a quantum dimension for the the $\mathbb{M}$-modules in the above pair $\left(V^{\mathbb{M}}_i,M_i\right)$. The definition \eqref{eq:character} does not directly apply to an $\mathbb{M}$-module, but one can extend the definition using the $n$-graded dimension of $\mathbb{M}$-modules $M_i$'s. We define $\text{ch}_q M_i$ as \footnote{We are deeply grateful to Richard E. Borcherds for suggesting this alternative formula. It is similar to the generating function of multiplicity $\textbf{m}_i(-1,n)$ in Section 8.6 of \cite{Duncan1}, but without normalization by $1/|\mathbb{M}|$. }
\begin{equation}
\label{eq:chq}
\text{ch}_q M_i\equiv\sum_{\sigma}j_\sigma\cdot\overline{\chi_i(\sigma)}.
\end{equation}

Here $j_{\sigma}\equiv\sum_{n=-1}^{\infty} \chi_{V_n^\natural}(\sigma)q^n$ is the monstrous McKay-Thompson series for each $\sigma$ as well as the unique Hauptmodul for a genus-0 subgroup $\Gamma_{\sigma}$ of $SL(2,\mathbb{R})$ for each $\sigma$ \cite{ConwayNorton,Borcherds}. $\sigma$ belongs to an index set with order 171, deduced from the 194 conjugacy classes of $\mathbb{M}$. The difference $194-171=23$ can be understood from the one-to-one correspondence between conjugacy classes and irreducible representations of $\mathbb{M}$: most of the $194$ irreducible representations have distinct dimensions, except for $23$ coincidences. $\sigma$'s are only sensitive to the dimensions of the corresponding irreducible representations. $\overline{\chi_i(\sigma)}$ is complex conjugation of the character of the irreducible representation $M_i$ of the 171 ``conjugacy class'' $\sigma$.\footnote{In literature this is often denoted by $\overline{\text{tr}(\sigma|M_i)}$ or $\overline{\text{tr}(M_i(\sigma))}$ or $\overline{\text{ch}_{M_i}(\sigma)}$ as well.} At large $n$, summation in $\text{ch}_qM_i$ is dominated by the first Hauptmodul for the identity of $\mathbb{M}$, which is exactly the Klein's invariant $j(q)$, so that
\begin{equation}
\lim_{q\rightarrow1^-}\text{ch}_qM_i\approx j(q) \times d_i.
\end{equation}
In other words, one can view ch$_qM_i$ as a function ch$_qM_i(g)$ on group $\mathbb{M}$, and when defining the quantum dimension in \eqref{eq:chq}, we take the value when its argument is the identity element.

With this, we can define the quantum dimension of $\mathbb{M}$-modules $M_i$ in \eqref{eq:decomposition} relative to $V^{\natural}$ as
\begin{equation}
\label{eq:qdim}
\text{qdim}_{V^\natural}M_i\equiv\text{lim}_{q\rightarrow 1^-}\frac{\text{ch}_q M_i}{\text{ch}_q V^\natural}=\lim_{n\rightarrow\infty} \frac{\dim (M_i)_n}{\dim V^{\natural}_n}.
\end{equation}
Here $\text{ch}_q V^\natural=J(q)$ by applying \eqref{eq:character} to $V^{\natural}$, which is a $V^{\natural}$-module of itself. Combining the discussions above, the quantum dimension is just
\begin{equation}
\text{qdim}_{V^\natural}M_i=d_i.
\end{equation}

The $d_i$'s that appeared explicitly in \eqref{eq:J} of the TEE calculation are quantum dimensions of $M_i$, while those in \eqref{eq:asymptotic} are quantum dimensions of $V^{\mathbb{M}}_i$. They coincide numerically. As we mentioned before, the rationality of $V^{\mathbb{M}}$ is widely conjectured to be true\footnote{Unfortunately, the conjecture has only been proved only when the subgroup of the automorphism group is solvable \cite{Carnahan, Miyamoto}, which is not our case.}, and by a theorem of Huang \cite{Huang}, the module category of any rational, $C_2$-cofinite VOA is modular, i.e. it is a modular tensor category with a non-degenerate $S$-matrix. If one believes in the rationality conjecture, then $\text{qdim}_{V^{\mathbb{M}}}V^{\mathbb{M}}_i$'s have a well-defined interpretation in terms of modular $S$-matrices of the orbifold CFT $V^{\mathbb{M}}$:
\begin{equation}
\label{eq:smatrix}
d_i=S_{i0}/S_{00}.
\end{equation}
Note that these 194 ``anyons'' are the pure charge exitations in the corresponding topological ordered system described by the modular tensor category associated with the orbifold VOA $V^\mathbb{M}$.

\section{Discussion and Outlook}
\label{sec:summary}

In the high-temperature regime, the full modular-invariant partition function \eqref{eq:sum} is dominated by the black hole solution $Z_{1,0}(\tau)$, while in the low-temperature regime, it is dominated by $Z_{0,1}(\tau)$, the thermal \ads solution \cite{Tail, MaloneyWitten}. It is widely believed that there exists a Hawking-Page \cite{HawkingPage,Daives} transition at the critical temperature $\beta\sim 1$, or $r_+\sim l$. However, there is no consensus on whether this transition really exists \cite{MaloneyWitten,EvenOdd,NoTransition}, or if it exists, whether it is a first-order or a continuous phase transition \cite{Caputa1, Eune, Cappiello, Kurita, Sokolowski, Stephens}, or something else that is more subtle.  In this section we offer a clue from the TEE perspective.

We compare the $a=1$ (defined in Fig. \ref{fig:a}) case in \eqref{eq:ads3} of thermal \ads and the Fig. \ref{fig:alternative} case of a single-sided black hole, for their subregion $A$'s both cover the whole space. One then observes that even at the tree level, TEE of BTZ and thermal \ads have different signs. A natural guess would thus be that, if the transition exists, it should be topological and happen at where the TEE changes sign.

Our definition of topological entanglement entropy is the constant subleading term in the expression for entanglement entropy, which is in general different from the tripartite information as used in \cite{KitaevPreskill}. For topological phases in condensed matter physics, these differ by a factor of two and are both negative. For gravitational theories in the bulk, our topological entanglement entropies can be either positive (as in BTZ black hole case) or negative (as in the thermal \ads case). To calculate the tripartite information, one can use the surgery method presented in this paper and find the time dependence, which at late times is negative of the Bekenstein-Hawking entropy \cite{Future}. This matches with the results in CFTs with gravitational dual, it is expected that the tripartite information should be negative \cite{Monogamy} and that for thermofield double state, it equals negative of the Bekenstein-Hawking entropy \cite{Channel}. 

Quantum dimensions also appears in the calculation of left-right entanglement in RCFT \cite{LeftRight}. One might perform similar computations in the orbifold VOA $V^{\mathbb{M}}$ appeared in section \ref{sec:qdim}, by using the Ishibashi boundary CFT states that were constructed in \cite{MonstrousBrane} for open bosonic strings ending on D-branes.

Given the anyonic interpretation in section \ref{sec:whole}, one natural question to ask is that, to what extent 3d pure quantum gravity can be described as a theory of topological order. Naively one would expect the corresponding topological order to be the 3d Dijkgraaf-Witten theory of the monster group $\mathbb{M}$, which gives rise to the same modular tensor category as the one given by orbifold CFT $V^{\mathbb{M}}$ as explained in section \ref{sec:qdim}. On the other hand, it is also natural to expect the corresponding topological order to be the one which is effectively described by the double $SL(2,\mathbb{C})$ Chern-Simons theory. It would be highly non-trivial to find a mechanism that reconciles these two theories.

Another remark is that we have specified the bipartitions to be done at $t=0$ in section \ref{sec:btz}, while in general the result can be time-dependent. In the latter case one can still use the surgery method proposed in this paper to find the TEE or R\'enyi entropies, which can serve as an indicator of scrambling \cite{scrambling}.

A final mathematically motivated direction is the following. Vaughn Jones considered how one von Neumann algebra can be embedded in another and developed subfactor theory \cite{Jones1}. In general, the Jones program is about how to embed one \emph{infinite} object into another, reminiscent of field extensions in abstract algebra, and quantum dimension is defined exactly in this spirit. It would be interesting to see how subfactor theory in general can help connect topological phases and pure quantum gravity \cite{Jones2}.


\acknowledgments
We are deeply grateful to Richard E. Borcherds for teaching us quantum dimensions of $\mathbb{M}$-modules over $V^{\natural}$. We appreciate Song He and Mudassir Moosa's suggestions on the manuscript, and thank Ori J. Ganor and Yong-Shi Wu for remarks on Hawking-Page transition. We thank Norihiro Iizuka and Seiji Terashima for explaining their work, Andreas W. W. Ludwig and Zhenghan Wang for extremely helpful comments on the moonshine module. We thank Diptarka Das, Shouvik Datta and Sridip Pal for explaining their work and pointing out Ref. \cite{MonstrousBrane} to us. Zhu-Xi thanks Herman Verlinde for comments on the sign of BTZ TEE, and Zheng-Cheng Gu, Muxin Han, Jian-dong Zhang for helpful discussions.
We also appreciate the workshop ``Mathematics of Topological Phases of Matter'' at SCGP, where part of the work was completed.

\appendix

\section{Bipartition for the Full Partition Function}
\label{app:partition}

In this appendix we justify that inputting $j$-invariant into the replica trick formula is a legal operation. We need to make sure that the horizon in the $SL(2,\mathbb{Z})$ family of Euclidean BTZ black holes is still at the central cord of their solid tori, so that we can cut along it. Although $j$-function contains contribution from thermal AdS$_3$ which contains no black holes, we will see later that this configuration contributes nothing at a higher enough finite temperature. For convenience we set $l=1$.

To see how Euclidean BTZ Schwarzschild coordinates transform under the $SL(2,\mathbb{Z})$ action on $\tau$, we need an intermediate FRW metric for the unexcited (before being quotiented by $\Gamma$) \ads with cylindrical topology, similar to the one mainly used in \cite{MaloneyWitten}:
\begin{equation}
\label{eq:FRW}
\begin{split}
ds^2&=\cosh^2\rho~d\Sigma^2+d\rho^2\\
&=-\sinh^2\rho(du-d\bar{u})^2+\cosh^2\rho(du+d\bar{u})^2+d\rho^2\\
&=\sinh^2\rho~d\phi^2+\cosh^2\rho~dt'^2+d\rho^2,
\end{split}
\end{equation}
where $2u\equiv i\phi-t$ and $2\bar{u}\equiv-i\phi-t$ parametrize the domain of discontinuity $\Sigma$, and $\rho$ indicates the radial direction. 

To obtain a Euclidean BTZ from this, we demand $2u\equiv(t-i\phi)/\tau'$, with $\tau'\equiv-1/\tau=\Phi+i\beta$ the modular parameter for BTZ black hole, and $\tau$ the modular parameter of thermal AdS$_3$. 
The identification in the BTZ spatial direction is automatic due to the periodicity in the $\mathbb{H}^3$ metric; $\text{Im}\tau'$ represents the time identification because it is the length of the time cycle, and $\text{Re}\tau'$ offers a spatial twist upon that identification, inducing an angular momentum by ``tilting'' the meridian.\footnote{Situation is almost identical in the thermal AdS$_3$ \eqref{eq:FRW}, where Im$\tau$ specifies the time identification, upon which Re$\tau$ indicates a spatial twist.}
Define the Schwarzschild radial coordinate $r$:
\begin{equation}
\sinh^2\rho=\frac{r^2-(\text{Im}(1/\tau'))^2}{|\tau'|^2},
\end{equation}
we obtain the Euclidean BTZ black hole in Schwarzschild coordinates for $r\geq\text{Im}(1/\tau')$:
\begin{equation}
ds^2=N^2dt^2+N(r)^{-2}{dr^2}+r^2[d\phi+N^{\phi}(r)dt]^2,
\end{equation}
where $N^2(r)={[r^2-(\text{Im}(1/\tau'))^2][r^2+(\text{Re}(1/\tau'))^2]}/{r^2}$, and $N^{\phi}(r)={(\text{Re}(1/\tau'))(\text{Im}(1/\tau'))}/{r^2}$. Now the outer horizon is at $r_+=\text{Im}(1/\tau')$. When an $SL(2,\mathbb{Z})$ transformation is applied $\tau'\rightarrow\tau''={1}/{(c\tau'+d)}={\tau}/{(d\tau-c)}$, $r$ becomes
\begin{equation}
\displaystyle{r''^2\rightarrow\frac{(c~\text{Re}\tau'+d)^2\sinh^2\rho+(c~\text{Im}\tau')^2\cosh^2\rho}{|c\tau'+d|^4}}.
\end{equation}

It is enough to just think of ${1}/{(c\tau'+d)}$ because there are only three independent parameters in $(a, b, c, d)$ due to the constraint $ad-bc=1$. One has the freedom to choose $a=0$, which fixes $-bc=1$, consequently ${(a\tau'+b)}/{(c\tau'+d)} ={-1}/{(c^2\tau'+cd)}$. Redefine $-c^2= c$ and $-cd= d$, then we arrive at ${1}/{(c\tau'+d)}$. The minus sign in both $c$ and $d$ is not a problem, because $(c,d)$ is equivalent to $(-c,-d)$.

Since $\sinh^2\rho=r^2\beta^2-1$, we have $\text{Im}\tau''=-{c\beta}/{(c^2\beta^2+d^2)}$, $\text{Re}\tau''={d}/{(c^2\beta^2+d^2)}$, implying a rotating black hole. Now we need to see if the new $r''$ is still at the horizon in the Schwarzschild coordinates associated to $\tau'$, and it suffices to check that $r''_+=\text{Im}\tau''$. This is indeed true. Hence no matter what $(c,d)$ we change into, as long as $\tau$ and $\tau''$ are $SL(2,\mathbb{Z})$ equivalent, $r''=r_+''\equiv\text{Im}\tau$ will be mapped to a segment on $z$-axis of spherical coordinate system for the upper half $\mathbb{H}^3$, so our cut is still valid.

\section{TEE from the Whole $J(q)$ Function}
\label{app:J}

Now we plug the entire $J$-function as the canonical partition function into \eqref{eq:replica}. We start from the definition of $j$-invariant $j(\tau)=J(\tau)-744\equiv E^3_4(\tau)/\Delta(\tau)$, where $\Delta=\eta^{24}(\tau)$ is the normalized modular discriminant. To find the derivative of $J(\tau)$, we make use of the Jacobi theta function $\displaystyle{\vartheta(f)\equiv f'-\frac{m}{12}E_2(\tau)f}$ \cite{Zagier}, where $E_j(\tau)$ is Eisenstein series of weight $j$ and $m$ is the weight of the arbitrary modular form $f$. Substituting $j(\tau)$ for $f$, we obtain
\begin{equation}
    \frac{d}{d\tau}j(\tau)=\vartheta(j(\tau))+E_2(\tau)j(\tau).
\end{equation}
We have made use of the fact that the weight of $j(\tau)$ is three times the weight of $E_4(\tau)$ by definition. One easily observes from the right hand side of above equation that the weight of $j(\tau)$ becomes $12+2=14$ after differentiation. Since the vector space of $SL(2,\mathbb{Z})$ modular forms of weight $14$ is spanned by $E^2_4(\tau)E_6(\tau)$ and has complex dimension $1$, we must have $\frac{d}{d\tau}j(\tau)\propto \frac{E_6(\tau)}{E_4(\tau)}j(\tau)$, up to a constant prefactor. This factor can be found from plugging in the first several terms of the $j(\tau)$ function and we finally arrive at\footnote{It is also a consequence of applying Ramanujan's identities on $E_2$, $E_4$ and $E_6$ \cite{Das}.}
\begin{equation}
    \frac{d}{d\tau}j(\tau)=-2\pi i\frac{E_6(\tau)}{E_4(\tau)}j(\tau).
\end{equation}
Plugging into the replica trick equation \eqref{eq:replica} we obtain for the holomorphic part
\begin{equation}
S_{\text{full}}(\tau)=\ln J(\tau)+2\pi\beta\frac{j(\tau)}{J(\tau)}\frac{E_6(\tau)}{E_4(\tau)}.
\end{equation}

To calculate the ration $E_6/E_4$, we use the asymptotic formula for the holomorphic Einstein series $G_{s}(\tau)\equiv2\zeta(s)E_{s}(\tau)$, assuming $0<|\arg \tau|<\pi$ and Re$(s)>-N+1$ for any positive integer $N$ \cite{Eisenstein}:
\begin{equation}
\begin{split}
G_s(\tau)&=(1+\tau^{-s})(1+e^{\pi is})\zeta(s)+2\sin (s\pi)\frac{\zeta(s-1)}{s-1}\tau^{-1}-\left(1+\cos (s\pi)\right)\zeta(s)\\
&\quad+\sum_{k=1,\,k\text{ odd}}^{N-1} 2\sin (s\pi) {-s \choose k}\zeta(s+k)\zeta(-k)\tau^k+\mathcal{O}(|\tau|^N),\quad |\tau|\leq1.
\end{split}
\end{equation}
For both $s=4,6$, the second term vanishes at high temperatures $|\tau|=\rightarrow0$, and $\sin (s\pi)$ in the summation over $k$ vanishes as well. Switching to the real variable $\beta=-i\tau$, we have $G_4(i\beta)\approx2\beta^4\zeta(4)$ and $G_6(i\beta)\approx-2\beta^6\zeta(6)$ as $\beta\rightarrow0$. And since in this limit, $j(i\beta)\approx J(i\beta)$, we have for $k=1$
\begin{equation}
S_{full}(\tau,\bar{\tau})\approx 2\ln J(\tau)-4\pi\beta^3,
\end{equation}
where we have taken into account the anti-holomorphic part. 

Now we see that if we consider the entire $SL(2,\mathbb{Z})$ family of black holes as well as thermal AdS$_3$ (the later contributes little at small $\beta$), the one-loop contribution to TEE is negative, agreeing with our previous calculations. 



\end{document}